\documentclass[prl, aps, amssymb, twocolumn, superscriptaddress, notitlepage, longbibliography]{revtex4-2}

\usepackage{bm,amssymb,amsmath,mathrsfs,amstext,latexsym,physics}
\usepackage{cancel,hhline}
\usepackage[usenames,dvipsnames]{xcolor}
\usepackage[colorlinks=true,citecolor=blue,urlcolor=blue,linkcolor=blue]{hyperref}

\usepackage[caption=false,subrefformat=parens,labelformat=empty]{subfig}
\usepackage{graphicx,tikz}
\usepackage{orcidlink}

\begin{document}
\title{Universal Growth of Krylov Complexity Across a Quantum Phase Transition}

\author{Andr\'as Grabarits\orcidlink{0000-0002-0633-7195}}
\email{andras.grabarits@uni.lu}
\affiliation{Department  of  Physics  and  Materials  Science,  University  of  Luxembourg,  L-1511  Luxembourg, Luxembourg}

\author{Adolfo~del Campo\orcidlink{0000-0003-2219-2851}}                 
\affiliation{Department  of  Physics  and  Materials  Science,  University  of  Luxembourg,  L-1511  Luxembourg,  Luxembourg}
\affiliation{Donostia International Physics Center,  E-20018 San Sebasti\'an, Spain}

\date{\today}


\begin{abstract}
{\bf We study the statistical properties of the spread complexity in the Krylov space of quantum systems driven across a quantum phase transition. Using the diabatic Magnus expansion, we map the evolution to an effective one-dimensional hopping model. For the transverse field Ising model, we establish an exact link between the growth of complexity and the Kibble-Zurek defect scaling: all cumulants of complexity exhibit the same power-law scaling as the defect density, with coefficients identical to the mean, and the full distribution asymptotically becomes Gaussian. We also provide a general scaling argument for the complexity growth across arbitrary second-order quantum phase transitions, which is further demonstrated numerically in the long-range Kitaev models, both for short and long-range couplings.}
\end{abstract}

\maketitle

\noindent{\textbf{\textsf{Introduction}}}

\noindent The growth of complexity in quantum many-body dynamics underlies fundamental questions of thermalization, chaos, and the computational hardness of describing quantum evolution. In time-independent settings, Krylov subspace methods have emerged as a versatile framework for quantifying the complexity of quantum evolution ~\cite{Viswanath1994,Nandy25,Baiguera2025,Rabinovici25}. By constructing a compact basis tailored to the dynamics, these methods capture essential features of operator spreading, entanglement growth, scrambling, and quantum chaos. Within the Krylov basis, the quantum time evolution maps onto an effective one-dimensional tight-binding model, where complexity is naturally identified with the average position of the evolving state. This notion of Krylov complexity has proven useful in a wide variety of scenarios, ranging from the study of operator growth and the spread of quantum complexity~\cite{Parker_2019,Barbon19,Rabinovici21,Caputa22,Caputa_QMChaos_2022,Hornedal2022,Caputa2023} to quantum control and variational quantum algorithms~\cite{Takahashi24,Visuri25}.

In contrast, the characterization of operator growth and Krylov complexity in time-dependent settings remains largely uncharted. The general framework proposed in Ref.~\cite{Takahashi25} employs a Floquet operator that is nonlocal in time, making its application challenging in many-body systems. This limitation is especially significant given the prevalence of nonequilibrium processes in experimental platforms, from quantum quenches and annealing protocols to optimal control. It is therefore important to develop a versatile framework that both captures the growth of complexity and reveals its connection to universal features of complex and critical quantum dynamics.

In this context, the Kibble--Zurek (KZ) mechanism provides a well-established universal framework for defect formation in systems driven slowly across second-order quantum phase transitions at a finite rate  $1/\tau$~\cite{Kibble76a,Kibble76b,Zurek85,Zurek96c,DZ14, Dziarmaga2005Dynamics}. The KZ mechanism predicts the breakdown of adiabaticity near criticality in the freeze-out timescale $\hat t=\left(\tau_0\tau^{z\nu}\right)^{\frac{1}{z\nu+1}}$. The associated correlation length  $\hat\xi=\xi_0(\tau/\tau_0)^{\frac{\nu}{z\nu+1}}$ sets the typical size of the domains in the broken-symmetry phase. This yields the KZ scaling for the average number of resulting topological defects. For instance, for $D$-dimensional defects in $d$ spatial dimensions, one finds the universal power law $n\propto\hat\xi^{-(d-D)}\sim\tau^{-\frac{(d-D)\nu}{z\nu+1}}$ \cite{DZ14}. Although the KZ mechanism has primarily been employed to quantify the defect number and related observables, an open challenge remains to determine whether other physical quantities, such as the complexity of quantum evolution, also exhibit nonequilibrium universality.

In this work, we introduce an exact and versatile analytical framework to describe the growth of Krylov complexity in time-dependent settings and to characterize the dynamics of quantum phase transitions (QPTs). Our formalism employs the diabatic Magnus expansion, e.g., relative to the instantaneous ground state. Using the transverse-field Ising model (TFIM) as a testbed, we show that all cumulants of Krylov complexity display universal power-law scaling governed by the KZ mechanism. Moreover, their statistics converge to a universal Gaussian distribution within the KZ regime. Building on these results, we establish general scaling arguments that provide a universal framework for quantifying the growth of complexity in quantum systems driven across critical points.

\noindent{\textbf{\textsf{Results}}}

\noindent\textbf{Krylov subspace from the diabatic Magnus operator.}
Consider a driven system with Hamiltonian $H(t)=\sum_nE_n(t)|n(t)\rangle \langle n(t)|$. The time evolution operator is given by the time-ordered exponential $U=\mathcal{T}\exp(-i\int_0^tH(s)ds)$, while the time-evolution operator associated with parallel transport explicitly reads $U_{\rm ad}=\sum_n|n(t)\rangle\langle n(0)|$ \cite{Berry09}. To capture the complexity arising exclusively from nonadiabatic contributions, we consider the diabatic time-evolution operator $\mathbb{U}(t)=U(t)U_{\rm ad}(t)^\dag$ that maps the instantaneous ground state to the time-evolved state at $t$. We define the diabatic Magnus operator as $\Omega(t)=i\,\log\left(\mathbb{U}(t)\right)$, in terms of which
    \begin{eqnarray}
    \lvert\psi(t)\rangle=e^{-i\,\Omega(t)}\lvert \mathrm{GS}(t)\rangle
    =\sum_{m=0}^\infty\,\frac{(-i)^m\Omega^m(t)}{m!}\lvert \mathrm{GS}(t)\rangle,\quad
    \end{eqnarray}
     where $\lvert\mathrm{GS}(t)\rangle$ is the instantaneous ground state. This construction avoids adiabatic contributions that would otherwise obscure the diabatic nonequilibrium effects. 
The corresponding Lanczos algorithm reads
\begin{eqnarray}
    &&\lvert K_{n+1,t}\rangle\,b_{n+1,t}=
    \Omega(t)\lvert K_{n,t}\rangle-a_{n,t}\lvert K_{n,t}\rangle-b_{n,t}\lvert K_{n-1,t}\rangle,\nonumber\\
    &&a_{n,t}=\langle K_{n,t}\lvert\Omega(t)\rvert K_{n,t}\rangle,\nonumber\\
    &&b_{n,t}=\langle K_{n-1,t}\lvert\Omega(t)\rvert K_{n,t}\rangle,\label{KrylOmega}
\end{eqnarray}
where the Krylov basis and the Lanczos coefficients at different times are generated independently, easing the implementation of the algorithm with respect to alternative constructions that are nonlocal in time \cite{Takahashi25}. 

The Krylov wave function is obtained from the occupation amplitudes given by the overlap of the time-evolved state with the Krylov basis elements, $\varphi_n(t)=\langle K_n(t)\vert \psi(t)\rangle$.
Choosing the first Krylov state as the instantaneous ground state, $\lvert K_0(t)\rangle=\lvert\mathrm{GS}(t)\rangle$, the Krylov wave function can be expressed by inserting the identity in terms of the Krylov states at time $t$,
\begin{equation}
\begin{split}
    \varphi_n(t)&=\sum_m\langle K_n(t)\lvert e^{-i\Omega(t)}\vert K_m(t)\rangle\langle K_m(t)\rvert K_0(t)\rangle\\
    &=\langle K_n(t)\lvert e^{-i\Omega(t)}\rvert K_0(t)\rangle.
    \end{split}
\end{equation}
Accordingly, the time-evolving state can be written in terms of the Krylov wave functions as $\lvert\psi(t)\rangle=\sum_n\lvert K_n(t)\rangle\varphi_n(t)$ and it follows from (\ref{KrylOmega}) that the diabatic Magnus operator admits a tridiagonal form in the Krylov basis. 

\noindent\textbf{Magnus operator in the transverse-field Ising model (TFIM).}
We first consider TFIM describing the spin chain with ferromagnetic interactions~\cite {Suzuki2012Quantum}, $\hat{H}(t) = - J \sum_{j=1}^L \left[ \hat{\sigma}^z_j \hat{\sigma}^z_{j+1} + g(t) \hat{\sigma}_j^x \right]$,
which provides an ideal testbed for studying second-order QPTs~\cite{Zurek2005Dynamics, Dziarmaga2005Dynamics,delcampo12,Adolfo2018_KinkStat, Bando20, King22, Balducci2023Large}. Its fermionic representation reduces to the sum of independent two-level systems (TLSs) given by the Hamiltonian
$\hat{H} = 2 \sum_{k} \hat{\psi}_k^\dagger \left[ (g(t)-\cos k)\tau^z + \sin k \, \tau^x \right] \hat{\psi}_k
    = 2 \sum_{k} \hat{\psi}_k^\dagger H_k(t) \hat{\psi}_k$,
where $\hat{\psi}_k := (\hat{c}_k, \hat{c}_{-k}^\dagger)^T$ contains the creation and annihilation operators for fermions of momentum $k=\pm\frac{\pi}{L},\,\pm\frac{3\pi}{L},\,\pm\pi\mp\frac{\pi}{L}$, and $\tau^{x,y,z}$ denote the Pauli matrices acting on each TLS. 
We consider linear driving protocols, $g(t) =  -t/\tau$, initialized at $t = -g_0 \tau$ in the paramagnetic phase, and terminating at $g(0)=0$. The finite-rate crossing of the QPT at $g_c=1$ induces topological defects as kink pairs, corresponding to excitations of the $(-k,k)$ doublets in the TLSs. For $k < \pi/2$, the single-particle levels $\epsilon_k(t)$ undergo avoided level crossings, with the minimal level separation given by $\sin k$ for the $k$-th mode. The KZ scaling of defects in the slow-driving regime, $\tau > 1$, is captured by a sum of independent Landau-Zener transition probabilities~\cite{Dziarmaga2005Dynamics,Damski05,DamskiZurek06},
$p_k \approx e^{-2\pi k^2 \tau}$, leading to
$n \approx \frac{1}{2\pi} \int_0^{\pi/2} \mathrm dk\, p_k \sim \tau^{-1/2}$,
in agreement with the original KZ prediction $n \sim \tau^{-d\nu/(1+z\nu)}$ in $d$ dimensions for point-like defects, with critical exponents $z = \nu = 1$~\cite{Zurek2005Dynamics}.

As the time-evolution operator is the product of unitaries over each $k$ mode $\mathbb{U}(t)=\prod_k\,\mathbb{U}_k(t)$,  the corresponding Magnus operator is given by the direct sum $\Omega(t)=\sum_k\Omega_k(t)$, with $\Omega_k(t)=-i\log(\mathbb{U}_k(t))$. The summation is understood so that the $k$-th term only acts on the corresponding TLS.
In addition, for $t/\tau\ll 1$ near the final state, the Magnus operator has matrix elements satisfying $\Omega^{21}_k(t,\tau)=\left(\Omega^{12}_k(t,\tau)\right)^*$,
    $\Omega^{22}_k(t,\tau)=-\Omega^{11}_k(t,\tau)$, with 
\begin{eqnarray}\label{eq:Magnus_k}
    &&\Omega^{11}_k(t,\tau)\approx\\
    &&\frac{\mathrm{arg}\left[\sqrt{q_k}\cos\Phi_k-i\sqrt{1-q_k\cos^2\Phi_k}\right]\sqrt{q_k}\sin\Phi_k}{\sqrt{1-q_k\cos^2\Phi_k}}\nonumber\\    
    &&\Omega^{12}_k(t,\tau)\approx\\
    &&i\frac{\mathrm{arg}\left[\sqrt{q_k}\cos\Phi_k+i\sqrt{1-q_k\cos^2\Phi_k}\right]\sqrt{p_k}e^{-i\omega_k}}{\sqrt{1-q_k\cos^2\Phi_k}}, \nonumber
    \end{eqnarray}
    where $q_k=1-p_k$, as shown in~\cite{supp}. Further, the angles admit the leading-order behavior, $\omega_k\approx3\pi/4-\tau(1+t/\tau)^2$, $\Phi_k\approx\tau(1+t/\tau)^2$~\cite{Moessner_KZM_2024}.
    
\noindent\textbf{Krylov-wave function and Lanczos coefficients.}
 For slow quenches across the quantum critical point, in the KZ scaling regime, the $n$-th Krylov state is given by the superposition of the $n$-times excited product states with their coefficients given by the corresponding products of the off-diagonal elements of the Magnus operator; see~a detailed derivation in \cite{supp}. This limit holds whenever $n\ll L\tau^{-1/2}$ for the number of Krylov states, above which the Krylov wave function vanishes in the leading order,
    \begin{eqnarray}\label{eq:Krylov_states}
        &&\lvert K_n\rangle\approx\mathcal N_n\\
        &&\times\sum_{k_1\neq\dots\neq k_n}\prod_{k_1\neq\dots\neq k_n}(\Omega^{12}_k,0)^T_k\prod_{k\notin\{k_1\neq\dots k_n\}}(0,1)^T_{k}.\nonumber
    \end{eqnarray}
    Here, $\mathcal N_n$ is the normalization factor of the $n$-th Krylov state.
    The normalization scales in the leading order as
    \begin{eqnarray}
        \mathcal N^{-2}_n\sim n!\left(L\tau^{-1/2}\right)^n,
    \end{eqnarray}
    where the integral approximation and the leading-order behavior of the non-equal multiple sum were exploited. The normalization can be expressed as the product of the off-diagonal Lanczos coefficients $\mathcal N^{-1}_n=\prod_{l=1}^n\,b_l$, which admit the following leading-order scaling behavior,
\begin{equation}\label{eq:a_n_b_n}
b_n\sim L^{1/2}\tau^{-1/4}\sqrt{n},\,a_n\sim L\tau^{-1/2}.
\end{equation}
The dependence of the off-diagonal coefficients on $n$ is shown for various values of $\tau$ and $L$ in Fig.~\ref{fig:a_n_b_n}, exhibiting a precise scaling collapse when rescaled with $\tau^{1/4}L^{-1/2}$.
    \begin{figure}
    \centering
    \includegraphics[width=\columnwidth]{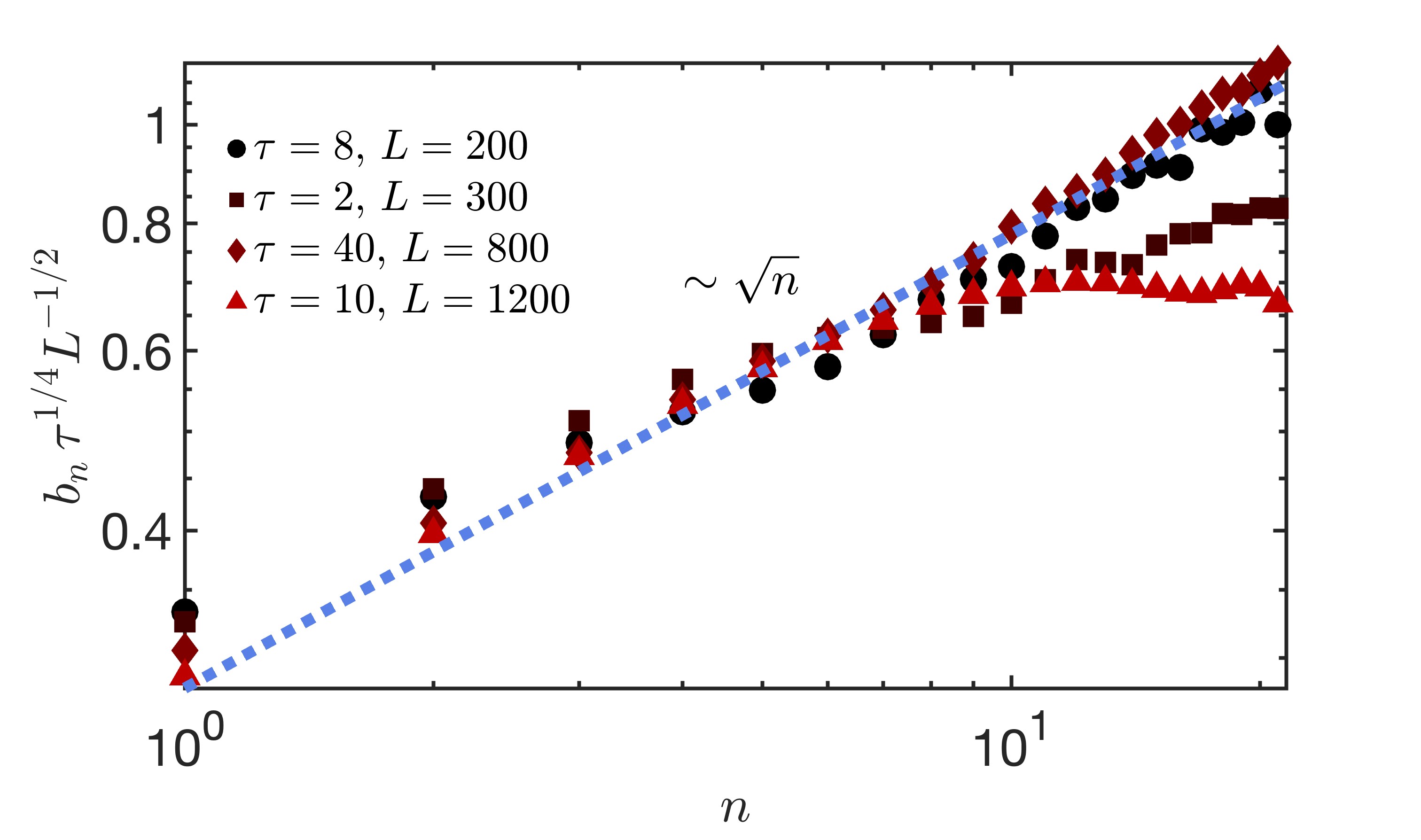}
    \caption{Off-diagonal Lanczos coefficients for different system sizes and driving times, all of them growing approximately as $\sim\sqrt n$, and collapsing onto a single curve by the proper rescaling given in Eq.~\eqref{eq:a_n_b_n}.}\label{fig:a_n_b_n}
\end{figure}

The corresponding Krylov wave function in the same leading-order approximation takes the form 
\begin{eqnarray}\label{eq: varphi_n}
    \varphi_n(t)\approx\frac{(2CL)^{n/2}\tau^{-n/4}}{\sqrt{n!}}e^{-CL\tau^{-1/2}}e^{i(L+n)\tau(1+t/\tau)^2-in\pi/2},\nonumber\\
\end{eqnarray}
that determines the Krylov complexity statistics, as discussed below.

As a caveat, the Magnus operator takes a direct-sum form rather than a product form, requiring the Krylov basis states to be constructed as tensor products of the $L/2$ two-level momentum states. This leads to exponential growth in computational cost, scaling as $\sim 2^{L/2}$, which limits exact numerical simulations to system sizes of up to $L \approx 48$. However, since only low-momentum modes contribute significantly in the KZ scaling regime, it is sufficient to truncate the system to an effective number of TLS, $L_\mathrm{eff}$, and run the Lanczos algorithm within this reduced basis, while still capturing all essential features of the full dynamics for $L > L_\mathrm{eff}$.

\emph{Krylov complexity statistics.---}
Spreading in the Krylov space is captured by the operator
    $\mathcal K=\sum_n\,n\lvert K_n(t)\rangle\langle K_n(t)\rvert$,
that weights the occupation linearly with the distance in the Krylov lattice.
The Krylov complexity is given by its average in the time-evolved state,
$K_1=\langle\psi(t)\lvert\mathcal K\rvert\psi(t)\rangle=\sum_n\,n\lvert\varphi_n(t)\rvert^2$.
    
A complete characterization of the spreading in Krylov space is provided by the probability distribution $P_{\mathcal K}(n)=\langle\psi(t)\lvert \delta[\mathcal K-n]\rvert\psi(t)\rangle=\sum_m\lvert\varphi_m(t)\rvert^2\delta[\mathcal K-m]$, shown in \ref{fig:Stat_Krylov}.
Qualitatively, the distribution is broad and shifts away from the origin at fast quench rates, when the distribution is well-described by Gaussian fits. By contrast, the distribution is asymmetric and peaks near the origin at the onset of adiabatic dynamics. 
    \begin{figure}
    \centering
    \includegraphics[width=\columnwidth]{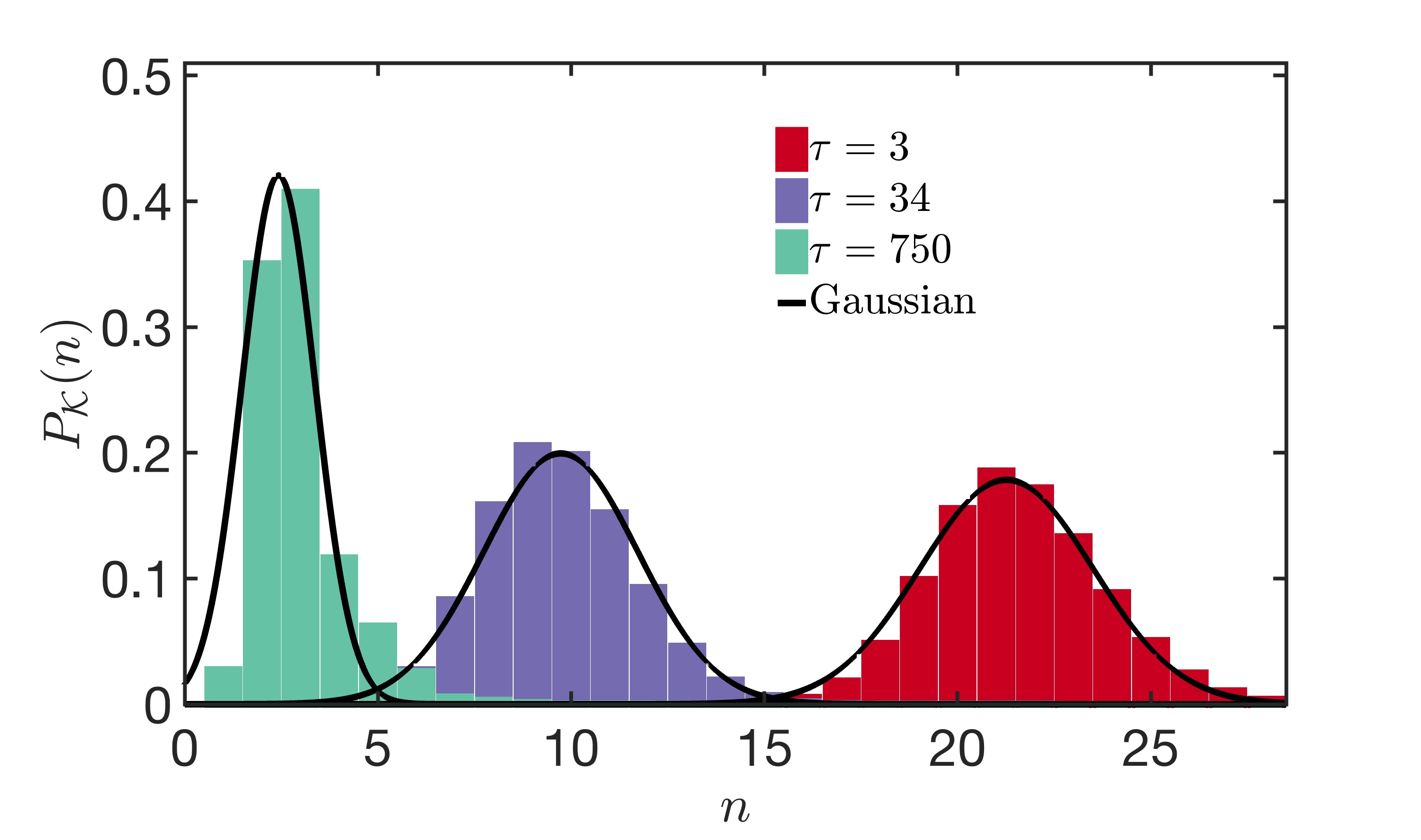}
    \caption{Statistics of Krylov complexity for a system of $L=400$ spins with an effective size set to $L_\mathrm{eff}=48$, capturing the low-energy spectrum that governs the dynamics. The solid lines show the agreement with the Gaussian approximation for $\tau=3$ and $34$ within the KZ scaling regime, while for $\tau=750$, slight deviations emerge as the system approaches the onset of adiabaticity.}\label{fig:Stat_Krylov}
\end{figure}
For a quantitative analysis, we consider    
the cumulant generating function,
    $\log\tilde P_{\mathcal K}\equiv\log\left[\langle\psi(t)\lvert e^{i\theta\mathcal K}\rvert\psi(t)\rangle\right]=\log\left[\sum_n\,\lvert\varphi_n(t)\rvert^2 e^{i\theta n}\right]$.
In terms of it,  higher-order cumulants of the distribution can be defined via the formal expansion $\log\tilde P_{\mathcal K}=\sum_{q=1}^\infty(i\theta)^qK_q/q!$. We focus on the first three cumulants. The mean $K_1$ equals the Krylov complexity, $K_2$  is the variance, and $K_3$ corresponds to the third centered moment.

Given the Krylov wavefunction in Eq.~\eqref{eq: varphi_n}, the cumulant generating function of the Krylov complexity simplifies to that of the Poisson distribution,
\begin{eqnarray}
    \log\tilde P_{\mathcal K}(\theta;\tau)\approx\frac{2(e^{i\theta}-1)CL}{\sqrt\tau},
\end{eqnarray}
where $C\approx 0.07 $ denotes a non-universal constant (see~\cite{supp}).
Thus, within this limit, the Krylov complexity cumulants become identical within the leading order of $L\tau^{-1/2}$
\begin{equation}\label{eq: cumulants}
    K_q\approx2CL\tau^{-1/2}.
\end{equation}
This result already evokes the proportionality between the defect cumulants~\cite {Adolfo2018_KinkStat}, but here the prefactors also become identical up to the leading order. The numerical confirmation of these scaling predictions is shown in Fig.~\ref{fig:cumulants} for the effective system size used in the Lanczos algorithm $L_\mathrm{eff}=46$. Cross-checked by the cumulants of the defect numbers, this size can precisely capture the dynamics of the low-energy subspace of $L=400$ up to the driving times $\tau \gtrsim 2$.
    \begin{figure}
    \centering
    \includegraphics[width=\columnwidth]{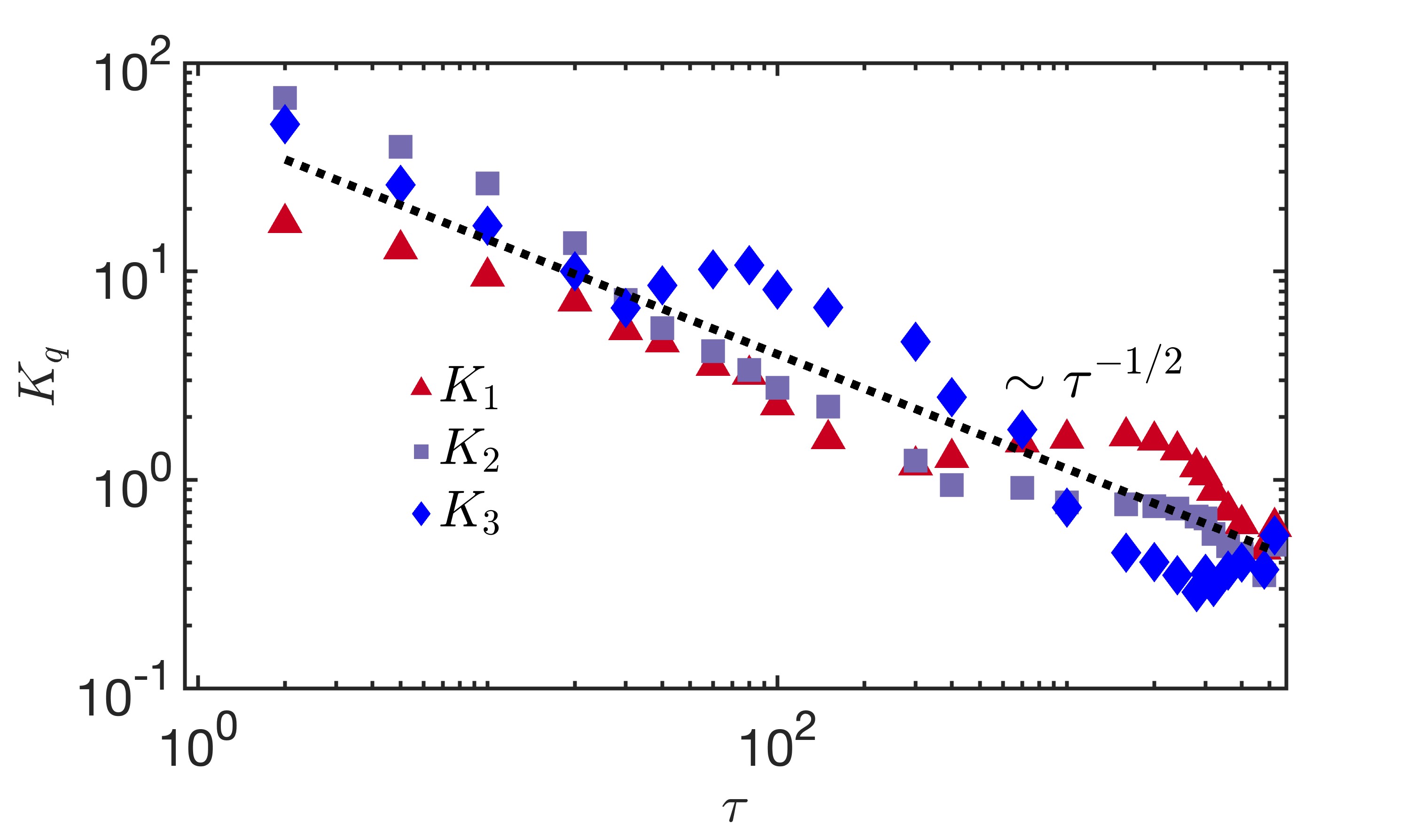}
    \caption{First three cumulants for the complexity as a function of $\tau$. With small fluctuations, they all follow the universal power laws predicted by Eq.~\eqref{eq: cumulants}, matching the KZ scaling of defect cumulants $(L=400,\,L_\mathrm{eff}=46)$.}\label{fig:cumulants}
\end{figure}

In the KZ scaling regime, the limiting distribution of Krylov complexity is indeed Gaussian, as the condition $L\tau^{-1/2} \gg 1$ guarantees that the central limit theorem applies to the underlying Poissonian statistics, 
\begin{equation}
     P_{\mathcal K}(n)\approx\frac{\exp\left[-\frac{(n-K_1)^2}{2K_2}\right]}{\sqrt{2\pi K_2}}.
\end{equation}
Alternatively, as all cumulants are proportional to this diverging parameter, all higher-order corrections beyond the variance become negligible.
This limiting behavior is presented in Fig.~\ref{fig:Stat_Krylov} for system sizes $L=400$ and with effective size $L_\mathrm{eff}=48$ for different driving times.

We also examine the time evolution of the Krylov complexity, i.e., under the generation of $\Omega(t,\tau)$ for arbitrary $g>0$ and for various final driving times. Near the final time ($t/\tau \ll 1$), the time dependence of the Magnus operator enters only through bounded functions, leaving the leading-order scaling of the complexity with respect to $\tau$, Eq.~\eqref{eq:Magnus_k}, unchanged. In addition to the overall KZ scaling, however, the complexity exhibits a pronounced oscillatory behavior that gradually diminishes as the system approaches the final point at $g=0$, smoothly reaching its asymptotic value. For times sufficiently far from $\tau$, within the symmetry-broken phase, these oscillations become increasingly pronounced and eventually dominate over the KZ scaling, signaling the onset of a nonuniversal dynamical regime governed by model-specific details. For $g>g_c$, the dynamics is approximately adiabatic, $\Omega(t)=0$, implying $K_q=0$ for all $q=1,2,3,\dots$ Near the critical point, however, a sharp increase in complexity occurs, indicating the presence of a second-order phase transition. Although this regime lies beyond the domain of validity of the leading-order approximation for the Krylov states in Eq.~\eqref{eq:Krylov_states} and the final-time analysis, the qualitative behavior and numerical results consistently capture the connection between defect formation and complexity growth. Remarkably, the Krylov complexity grows universally as a function of $g$, independent of $\tau$, in close agreement with the corresponding evolution of the excitation density. These features are illustrated in Fig.~\ref{fig:K_time} for various driving times within the KZ scaling regime.
\begin{figure}
    \centering
    \includegraphics[width=\columnwidth]{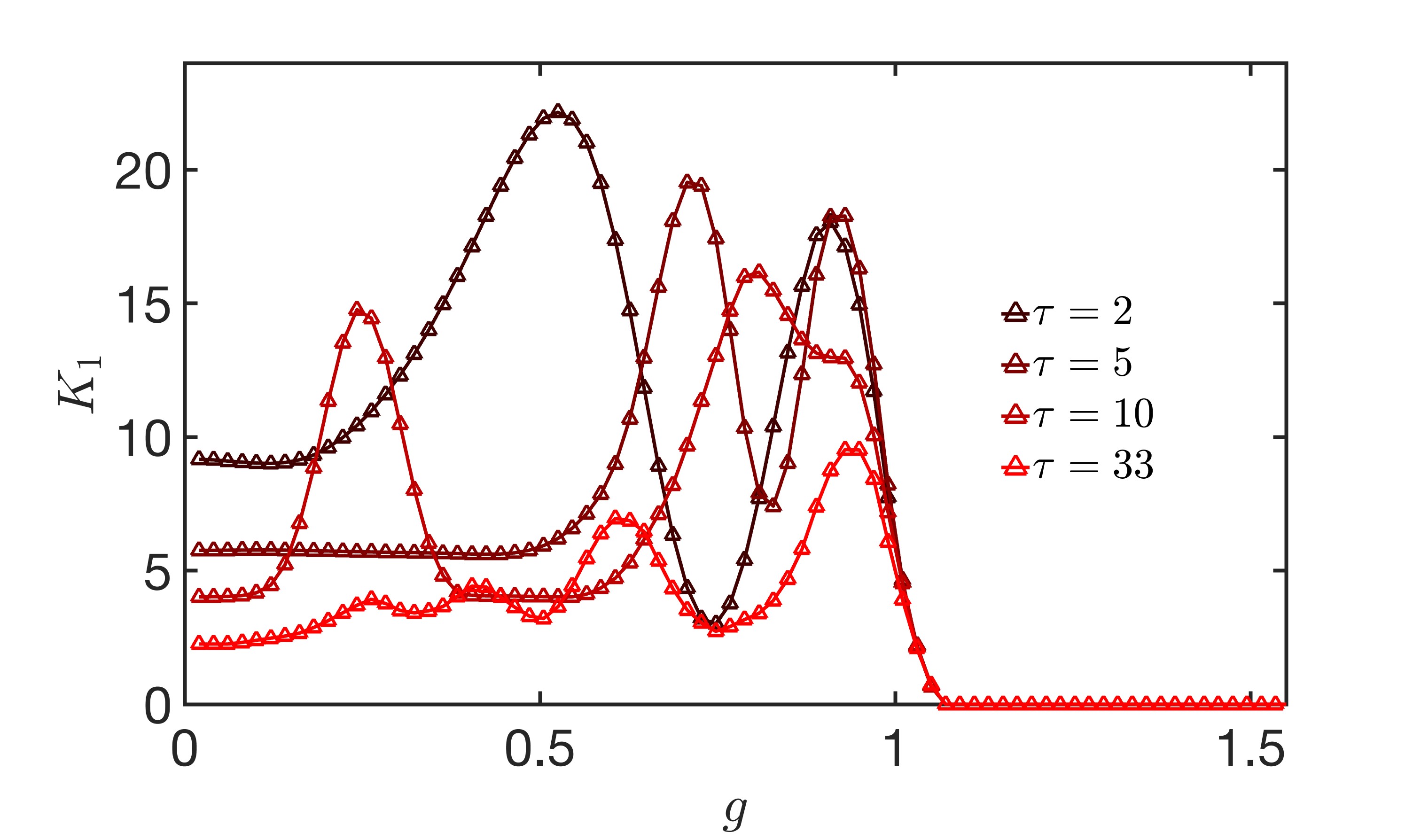}
    \caption{Time evolution of the Krylov complexity for different quench times. The sharp increase around $g_c=1$ captures the interplay between the QPT and complexity growth, followed by a non-universal and oscillatory regime that smoothly converges to the final value, as described by universal power laws. $(L=200,\,L_\mathrm{eff}=46)$.}\label{fig:K_time}
\end{figure}
These additional universal features around the critical point, as well as the highly oscillatory behavior at intermediate times, are also present in the variance and skewness of the complexity, as detailed in~\cite{supp}.

\noindent\textbf{Generalization to arbitrary critical systems.}
The negligible dependence on non-universal properties in TFIM, detailed in \cite{supp},  motivates the extension of the approach to a broader class of critical systems, anticipating similar robustness against model-specific details. For this, we consider $d$-dimensional uniform quantum critical systems described by a Hamiltonian of the free-fermionic form
$\hat H=\sum_{\mathbf k}\mathcal H_{\mathbf k}$, supporting a $(d-D)$-dimensional gapless critical surface ~\cite{Polkovnikov2005, Mondal_Sengupta_extendedKitaevKZM, Sarkar_Mandal_3DKitaevKZM, ZHANG_QUAN_CritSurfKZM_2022}. In these systems, the excitation amplitudes for slow drivings admit a universal leading-order contribution, $\langle\mathrm{ES}_{\mathbf{k}}\vert\psi_{\mathbf{k}}(\tau)\rangle\approx \sqrt{p\left(\mathbf k\tau^\alpha\right)}e^{i\theta_k}$ Consequently, the density of $D$-dimensional excitations is given by $n\approx\int\mathrm d\mathbf k^{d-D}\, p\left(\mathbf k\tau^\alpha\right)\sim\tau^{-(d-D)\alpha}$, which subsumes the KZ scaling and its extensions with modified power-law behavior, e.g., in long-range systems. In this universal power-law regime, the off-diagonal elements of the Magnus operator will also admit the same universal dependence on $\Omega^{12}_{\mathbf k}(\tau)\sim\Omega(\mathbf k\tau^\alpha)$ under the condition $\lvert\mathbf k\rvert\ll L^{d-D}\tau^{-\alpha}$ and for momenta perpendicular to the critical surface. For momentum states parallel to the critical surface, transitions are negligible, and the Magnus operator vanishes in leading order.

A similar analysis to that in the TFIM  leads to the conclusion that the $n$-th Krylov state equals the superposition of the $n$-times excited product states in the leading order for the same condition as above, $n\ll L^{d-D}\tau^{-\alpha(d-D)}$. Thus, both Lanczos coefficients and the Krylov occupation probability are expected to inherit the same behavior,
\begin{eqnarray}\label{eq:b_n_varphi_gen}
    &&b_{n}\sim L^{\frac{d-D}{2}}\sqrt n\tau^{-\frac{\alpha(d-D)}{2}},\, a_{n}\sim L^{d-D}\tau^{-\alpha(d-D)},\\
    &&\lvert\varphi_n(\tau)\rvert^2\sim \frac{\left(2CL^{d-D}\tau^{-\alpha(d-D)}\right)^n\,e^{-2C L^{d-D}\tau^{-\alpha(d-D)}}}{n!},
\end{eqnarray}
where $C\approx\int\mathrm d^{d-D}\mathbf k \lvert a(\mathbf k)\rvert^2$ is a generally non-universal constant depending on the functional form of the excitation probabilities along the orthogonal directions perpendicular to the critical surface, $\lvert a(\mathbf k)\rvert^2$, in the near adiabatic dynamics. The form of $\lvert\varphi_n(\tau)\rvert^2$ confirmed the Poissonian Krylov complexity statistics. 
As a result, the corresponding cumulants follow the same universal power laws,
\begin{eqnarray}\label{eq:K_q_gen}
    K_q\sim2CL^{d-D}\tau^{-\alpha(d-D)}.
\end{eqnarray}
More details of the general argument are presented in~\cite{supp}. 

\noindent\textbf{Demonstration in long-range Kitaev models.}
\begin{figure}[t]
    \centering    \includegraphics[width=0.49\textwidth,trim={0 11cm 0cm 0cm},clip]{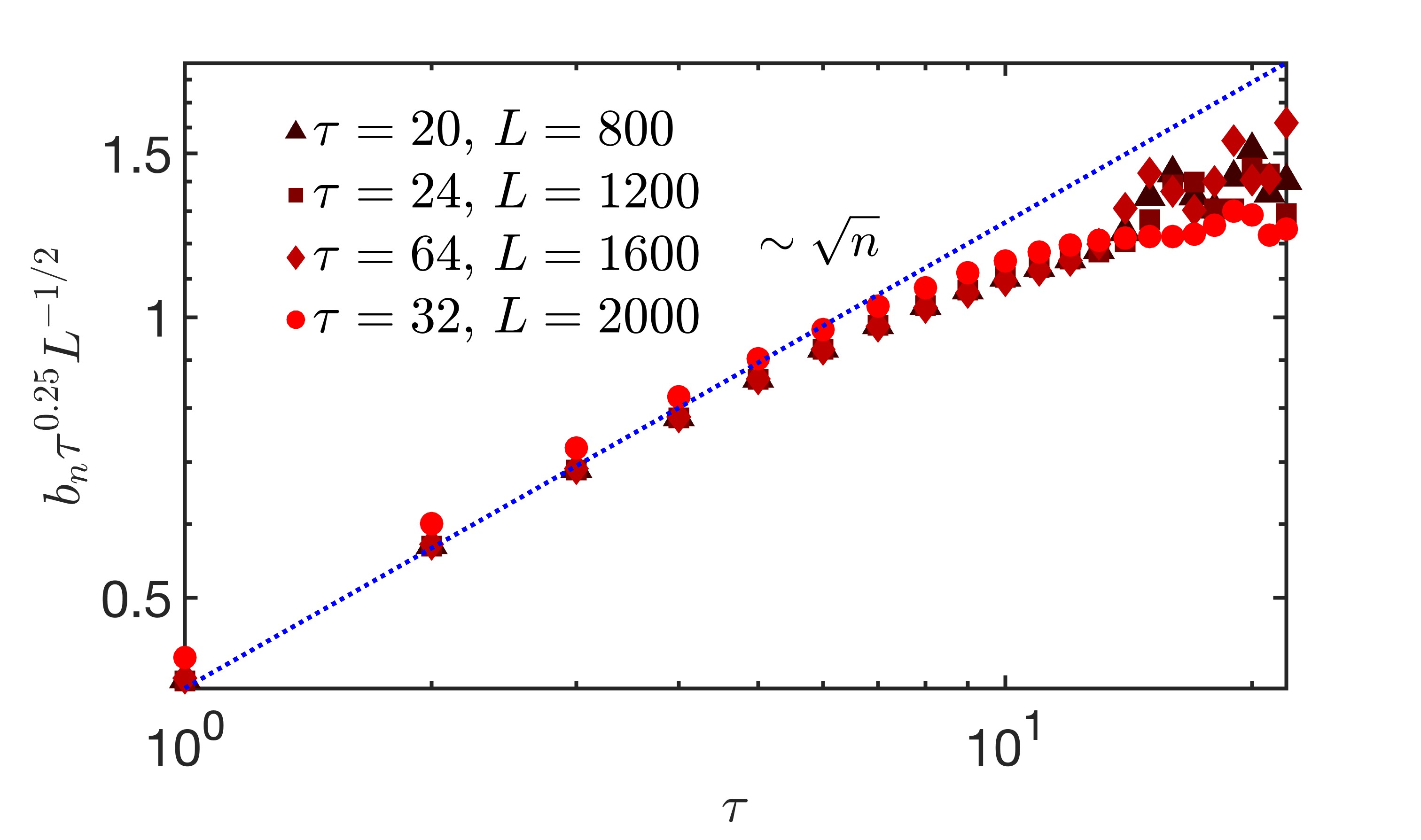}       \includegraphics[width=0.49\textwidth,trim={0 0  0 11 cm},clip]{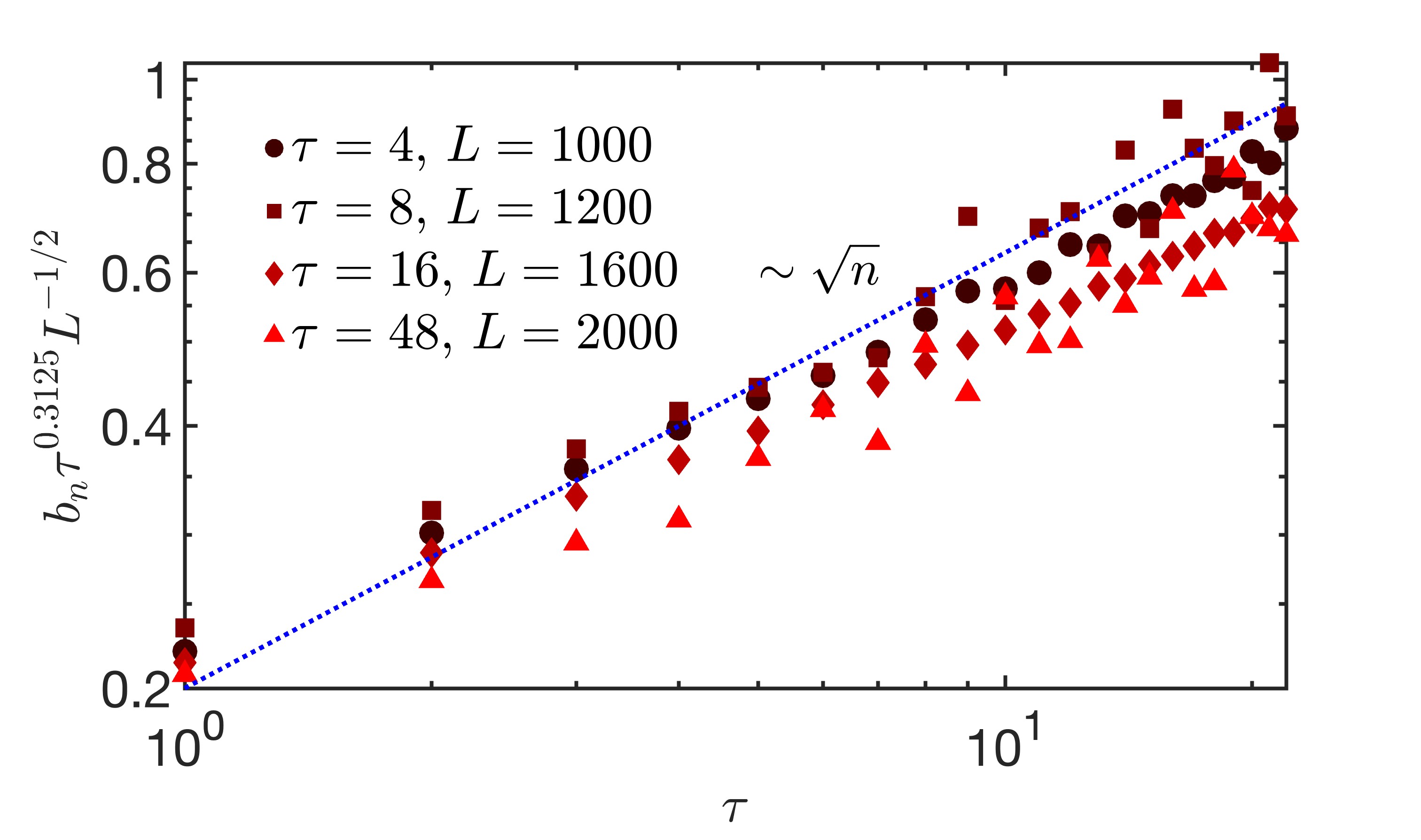}
    \caption{Lanczos coefficients in the LRKM. (a): Long-range hopping and short-range pairing regime, $\gamma=2.6,\,\beta=5$ with the coefficients showing precise collapse with the corresponding rescaling, growing universally with $\sim\sqrt n$,  up to good precision. (b): Long-range, dynamical scaling regime exhibiting similar universal growth of $\sqrt n$ with the corresponding rescalings with $\tau$}\label{fig:Krylov_LRKM_b_n}
    \end{figure}
Finally, we demonstrate the validity of our results by considering LRKMs describing spinless fermions with long-range $p$-wave hopping and pairing interactions on a one-dimensional lattice,
\begin{eqnarray}
\hat H\!=\!-\sum_i\sum_{r>0}\!\!\left(j_{r,\gamma} c_i^\dagger c_{i+r}\!+\!\Delta_{r,\beta} c_i^\dagger c_{i+r}^\dagger\right)\!+\!\mu\, c_i^\dagger c_i
+\mathrm{h.c.}\quad\,\,\,\,\,\,
\end{eqnarray}
The hopping and pairing amplitudes decay algebraically, 
$j_{r,\gamma}=J/N_\gamma\, r^{-\gamma}$ and $\Delta_{r,\beta}=d/N_\beta\, r^{-\beta}$, 
with exponents $\gamma,\beta>1$ and normalization factors $N_{\gamma,\beta}=2\sum_{r=1}^{N/2}r^{-\gamma,\,-\beta}$ ensuring extensivity and with $J=d=1$ taken for simplicity.
With these exponents, the LRKM exhibits a second-order QPT at $\mu_c=2$~\cite{Kitaev_Majorona_LR_2001}, which is robust against variations of the long-range exponents $\gamma,\beta>1$~\cite{Vodola_LRK, Dutta_LRK, LongRangePowerLawSC_Delgado_2017, LRK_powerlawDelAnna2017, Defenu_LRKM}. The resulting excitation density for slow drivings are fully governed by the pairing term, $\langle\hat N\rangle\propto \tau^{-\frac{1}{2(\beta-1)}}_Q,\,\beta <2,\,$ and $\langle\hat N\rangle\propto \tau^{-1/2}_Q,\,\beta >2,\,$ which violate the KZ scaling prediction in the dynamical scaling regime, for $\gamma<\beta,\,\gamma<2$. To this end, we test the general predictions in Eq.~\eqref{eq:b_n_varphi_gen} and Eq.~\eqref{eq:K_q_gen} both in the short- and long-range pairing regimes for long-range hopping interactions $\gamma=2.6,\,\beta=5$ and $\gamma=1.6,\,\beta=1.8$, respectively. 
\begin{figure}[h]
    \centering    \includegraphics[width=0.49\textwidth,trim={0 8.5cm 0cm 0cm},clip]{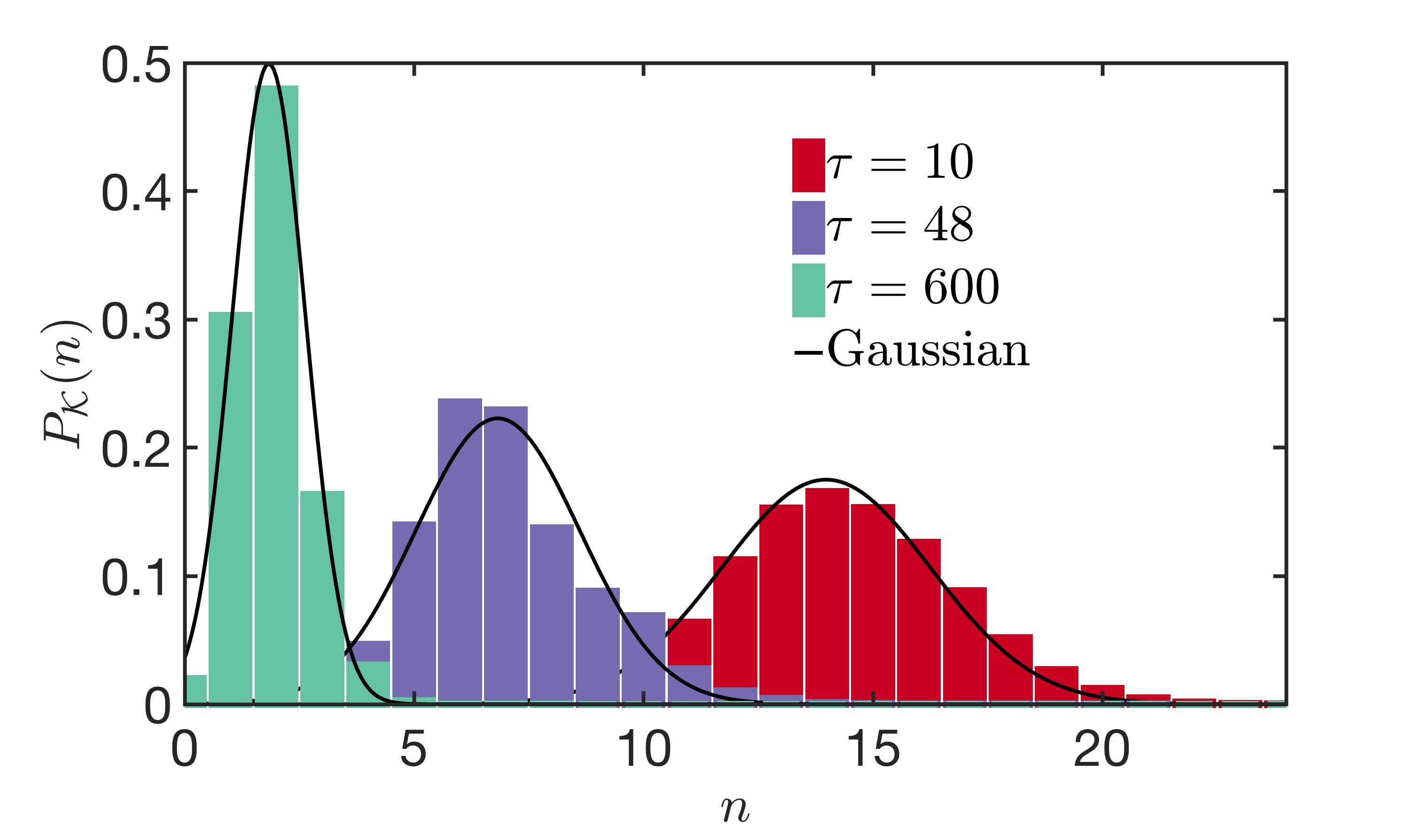}       \includegraphics[width=0.49\textwidth,trim={0 0  0 11 cm},clip]{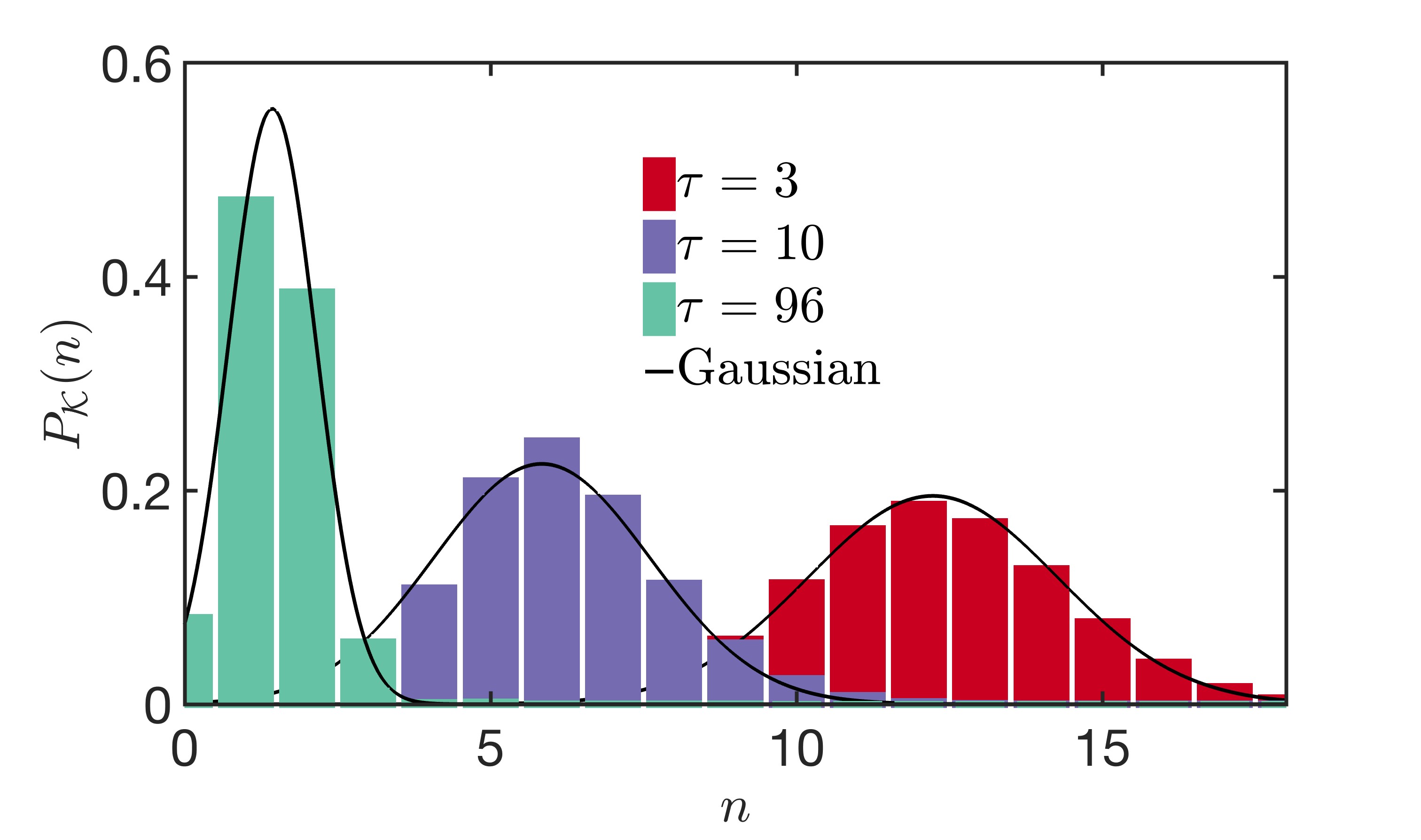}
    \caption{Histograms of complexities in the LRKM. (a): Long-range hopping and short-range pairing regime, $\gamma=2.6,\,\beta=5$ with $\tau=10,\,48,\,600$ with the distributions following approximately a Gaussian shape with their center and width decreasing gradually with increasing $\tau$. (b): Long-range, dynamical scaling regime with $\tau=3,\,10,\,96$ showing similar features.}\label{fig: Krylov_LRKM_Stat}
    \end{figure}
First, we show the results for the off-diagonal Lanczos coefficients in Fig.~\ref{fig:Krylov_LRKM_b_n}. In both cases, the predicted universal scaling with $\tau$ is reproduced up to good precision, exhibiting $\sim\sqrt n$ growth independently of the strengths of the hopping and pairing terms.
Next, we demonstrate the Gaussianity of the full distribution of the complexities in Fig.~\ref{fig: Krylov_LRKM_Stat}. In both long and short-range cases, histograms are consistent with Gaussian statistics, gradually shifted towards the origin with decreasing widths.
Finally, we also show that the first three complexity cumulants follow the predicted power laws up to good precision. As demonstrated in Fig.~\ref{fig: Krylov_LRKM_cumulants} in both cases, the first three cumulants are consistent with the predicted power-laws $K_q\sim\tau^{-1/2}$ for short-range pairings and $K_q\sim \tau^{-0.625}$ for long-range pairings.

\begin{figure}[h]
    \centering    \includegraphics[width=0.49\textwidth,trim={0 11cm 0cm 0cm},clip]{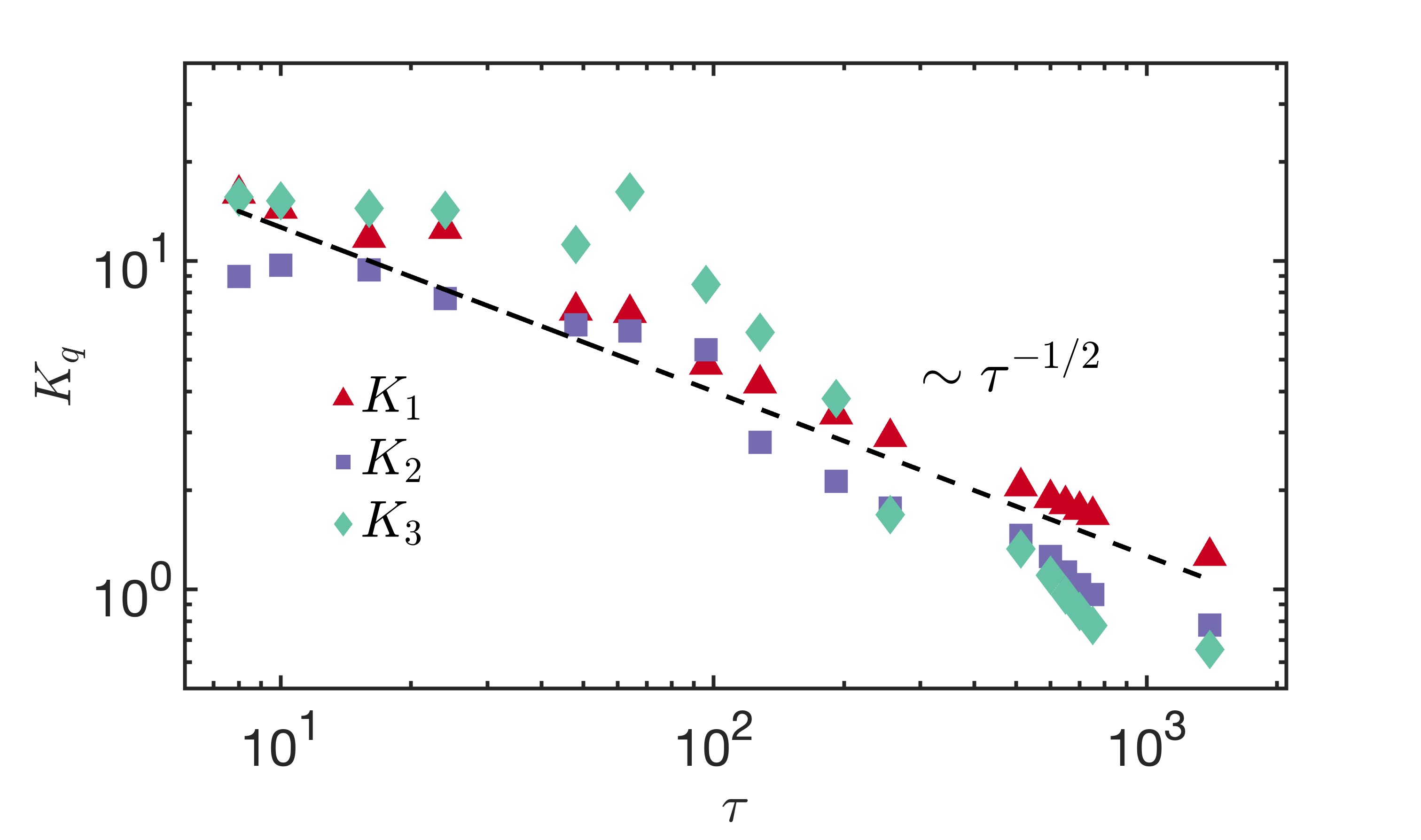}       \includegraphics[width=0.49\textwidth,trim={0 0  0 11 cm},clip]{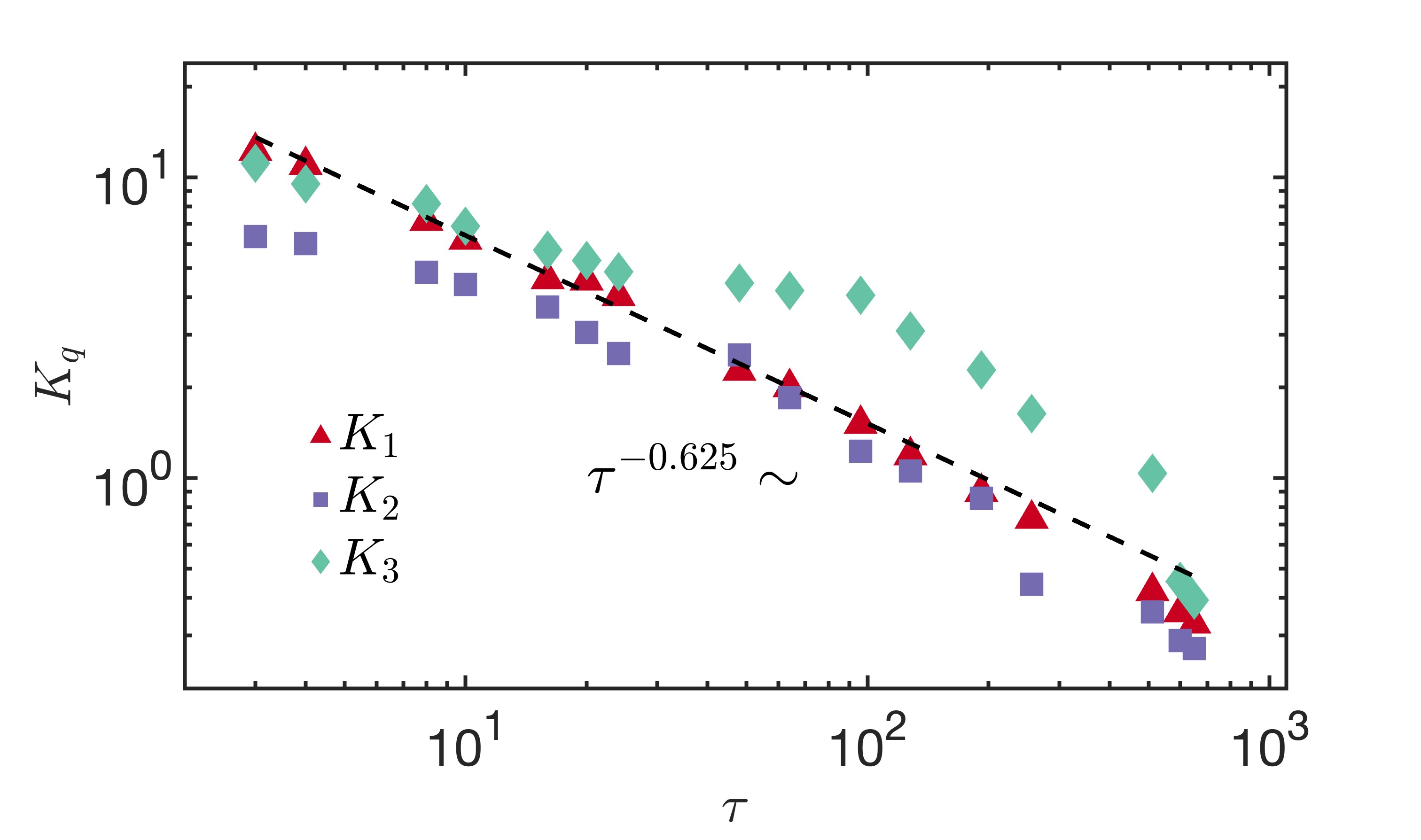}
    \caption{Complexity cumulants in the LRKM. (a): Long-range hopping and short-range pairing regime, $\gamma=2.6,\,\beta=5$ the cumulants follow the predicted KZ scaling law, $\sim\tau^{-1/2}$,  up to good precision. (b): Long-range, dynamical scaling regime with cumulants following the predicted $\tau^{-0.625}$ power-law.}\label{fig: Krylov_LRKM_cumulants}
    \end{figure}

\noindent{\textbf{\textsf{Discussion}}}

We have introduced a Krylov subspace approach for driven quantum systems based on the diabatic Magnus operator. Using this framework, we established the universal character of Krylov complexity dynamics during the slow crossing of quantum phase transitions. For the transverse-field Ising model, we obtained exact Lanczos coefficients and a Poissonian Krylov wavefunction, whose complexity cumulants coincide at leading order and exhibit Kibble–Zurek scaling, while the full statistics approach a Gaussian distribution. Extending these results, we proposed a general scaling argument applicable to other critical systems. These predictions were further demonstrated by the LRKMs with long-range hoppings for both long- and short-range pairings. In both cases, the complexity cumulants, statistics, and Lanczos coefficients matched the general predictions with good precision.

Our results thus reveal universal connections between quantum complexity and nonequilibrium phenomena that are accessible to experimental verification. 

\noindent{\textbf{\textsf{Methods}}}

\noindent\textbf{Krylov complexity statistics.}\label{app: Krylov_varphi}
To obtain the full statistics of the Krylov complexity, we first compute the Krylov wavefunction.  In the final basis, and by the leading order form of $U_k$ in the KZ scaling regime (see~\cite{supp}), the time-evolved wave-function becomes
\begin{eqnarray}
\lvert\psi(\tau)\rangle&=&\prod_k(\sin\frac{k}{2}u_k+\cos\frac{k}{2}v_k,-\cos\frac{k}{2}u_k+\sin\frac{k}{2}v_k)^T\,\\
    &\approx&\prod_k\left(v_k,-u_k\right)^T_k=\prod_k\left(\sqrt{p_k}e^{i\omega_k},-\sqrt{1-p_k}e^{i\Phi_k}\right)^T_k,\nonumber
\end{eqnarray}
 with the phases given in leading order as
 $\Phi_k=\tau(\cos k+t/\tau)^2+\tau\sin^2k\log\left[2\sqrt\tau\left(\cos k+t/\tau\right)\right]-\mathrm{arg}\Gamma(1+i\tau\sin^2k)\approx \tau(\cos k+t/\tau)^2+\tau\sin^2k\log\left[2\sqrt\tau\left(\cos k+t/\tau\right)\right]+\gamma_E\tau \sin^2k$ and $\omega_k=3\pi/4-\tau(\cos k+t/\tau)^2-\tau\sin^2k\log\left(\sqrt\tau(\cos k+t/\tau)\right]$~\cite{Moessner_KZM_2024, Nowak_Dziarmaga_2022}.

Using the leading order expression for the $n$-th Krylov basis vector,~\eqref{eq:Krylov_states} the corresponding Krylov wave function component reads 
\begin{widetext}
 \begin{eqnarray}
     \varphi_n(t,\tau)&\approx&
     (-1)^{L-n}\mathcal N_n\sum_{k_1\neq\dots\neq k_n}\prod_{l=1}^n\Omega^{21}_{k_l}\sqrt{p_{k_l}}e^{i\omega_{k_l}}\prod_{k\notin\{k_1,\dots,k_n\}}\sqrt{1-p_{k}}e^{i\Phi_{k}}\\     &=&(-1)^L(-i)^n\mathcal N_n\sum_{k_1\neq\dots\neq k_n}\prod_{l=1}^n\frac{\mathrm{arg}\left[\sqrt{1-p_k}\cos\Phi_k+i\sqrt{1-(1-p_k)\cos^2\Phi_k}\right]}{\sqrt{1-(1-p_{k_l})\cos^2\Phi_{k_l}}}\,\frac{p_{k_l}e^{-
     i\Phi_{k_l}}}{\sqrt{1-p_{k_l}}}\prod_{k}\sqrt{1-p_k}e^{i\Phi_k},\nonumber
 \end{eqnarray}
 where we have used the result for the off-diagonal matrix element of the Magnus operator, $$\Omega^{12}_k=i\frac{\mathrm{arg}\left[\sqrt{1-p_k}\cos\Phi_k+i\sqrt{1-(1-p_k)\cos^2\Phi_k}\right]}{\sqrt{1-(1-p_k)\cos^2\Phi_k}}\sqrt{p_k}e^{-i\omega_k},$$ (for further details see~\cite{supp}). 
 Using the leading-order behavior of the phases~\cite{supp}.
 The product of the normalizations is given by 
 \begin{eqnarray}
 \prod_k\left(1-e^{-2\pi\tau k^2}\right)^{1/2}e^{i\Phi_k}&\approx&e^{\frac{1}{2}\sum_k\log(1-e^{-2\pi\tau k^2})+iL\tau(1+t/\tau)^2}\\
 &&\approx e^{\frac{L}{4\pi}\int_0^\pi\mathrm dk \log\left[1-e^{-2\pi\tau k^2}\right]+iL\tau(1+t/\tau)^2}= e^{-C L\tau^{-1/2}+iL\tau(1+t/\tau)^2},\nonumber
 \end{eqnarray} 
 with $C\approx\zeta(3/2)/\sqrt{128\pi^2}$. Furthermore, the other phase factor in the first product gives $e^{in\tau(1+t/\tau)^2}$. The remaining part can be computed in the leading order as
 \begin{eqnarray}
     \sum_{k_1\neq\dots\neq k_n}\prod_{l=1}^n\frac{\mathrm{arg}\left[\sqrt{1-p_{k_l}}\cos\Phi_{k_l}+i\sqrt{1-(1-p_{k_l})\cos^2\Phi_{k_l}}\right]}{\sqrt{1-(1-p_{k_l})\cos^2\Phi_{k_l}}}\,\frac{p_{k_l}}{\sqrt{1-p_{k_l}}}\approx \left(\frac{L}{\sqrt{8\pi^3}}I(t,\tau)\right)^n\tau^{-n/2},
 \end{eqnarray}
 with the bounded integral function defined as
 \begin{eqnarray}
     I(t,\tau)=\int_0^\infty\mathrm dx\,\frac{\mathrm{arg}\left[\sqrt{1-e^{-x^2}}\cos\left[\tau(1+t/\tau)^2\right]+i\sqrt{1-(1-e^{-x^2})\cos^2\left[\tau(1+t/\tau)^2\right])}\right]}{\sqrt{1-(1-e^{-x^2})\cos^2\left[\tau(1+t/\tau)^2\right]}}\,\frac{e^{-x^2}}{\sqrt{1-e^{-x^2}}}.
 \end{eqnarray}
 Finally, the normalization is given by
 \begin{eqnarray}
     \mathcal N^{-2}_n=n!\sum_{k_1\neq\dots\neq k_n}\prod_{l=1}^n\lvert\Omega^{12}_{k_l}\rvert^2\approx \frac{L}{2\pi}\int_0^\pi\mathrm dk \left\lvert\Omega^{12}_k(t,\tau)\right\rvert^2\equiv n!\left(\frac{L}{\sqrt{8\pi^3}}I_2(t,\tau)\right)^n\tau^{-n/2},
 \end{eqnarray}
 where the factorial emerged as a combinatorial factor for counting the number of ways the excited state can be matched in the inner product.
 Here as well, a similar integral function has been introduced,
 \begin{eqnarray}
     I_2(t,\tau)=\int_0^\infty\mathrm dx\,\frac{\mathrm{arg}\left[\sqrt{1-e^{-x^2}}\cos\left[\tau(1+t/\tau)^2\right]+i\sqrt{1-(1-e^{-x^2})\cos^2\left[\tau(1+t/\tau)^2\right])}\right]^2}{1-(1-e^{-x^2})\cos^2\left[\tau(1+t/\tau)^2\right]}\,\frac{e^{-2x^2}}{1-e^{-x^2}}.
 \end{eqnarray}
 Although these two integral functions appear quite complicated, in the leading order they should respect the normalization of the Krylov wavefunction up to a phase factor that has already been fixed by $e^{i(n+L)\tau(1+t/\tau)^2}$. Reassuringly, the ratio of the two integral functions $I(t,\tau)/\sqrt{I_2(t,\tau)}$ is a slowly changing function within the interval $[0.112,0.139]$. This matches approximately the normalization factor $2C\approx 0.146$, while $I^2(t,\tau)/\sqrt{I_2(t,\tau)}/\sqrt{8\pi^3}\in[0.08,0.123]$.
 Thus, we neglect these contributions by restoring the overall normalization factor while keeping track of the additional time dependencies by the phase factor, 
 \begin{eqnarray}
     \varphi_n(t,\tau)&\approx& (-i)^n\left(\frac{L\,I^2(t,\tau)}{I_2(t,\tau)\sqrt{8\pi^3}}\right)^{n/2}\frac{\tau^{-n/4}}{\sqrt{n!}}\,e^{-CL\tau^{-1/2}}e^{i(L+n)\tau(1+t/\tau)^2}\\
     &\approx& \left(2CL\right)^{n/2}\frac{\tau^{-n/4}}{\sqrt{n!}}\,e^{-CL\tau^{-1/2}}e^{i(L+n)\tau(1+t/\tau)^2-in\,\pi/2}.\nonumber
 \end{eqnarray}
 \end{widetext}
Finally, the occupation probability reads 
 \begin{eqnarray}
     \lvert\varphi_n(\tau)\rvert^2\approx \frac{(2CL\tau^{-1/2})^n}{n!}e^{-2CL\tau^{-1/2}},
 \end{eqnarray}
 obeying a Poissonian probability law. This naturally leads to identical cumulants as stated in the main text,
 \begin{eqnarray}
     K_q=K_1\equiv\sum_n n\lvert\varphi_n(\tau)\rvert^2\approx 2CL\tau^{-1/2}.
 \end{eqnarray}

\noindent\textbf{Cumulants and distribution of Krylov complexity at intermediate times in the TFIM.}
We provide further details on the time-evolution of the Krylov complexity cumulants and the instantaneous complexity distributions in the TFIM. As shown in Fig.~\ref{fig:K_Stat}, for transverse fields in the paramagnetic phase, a negligible Krylov complexity spread is observable, as mirrored by its peaked distribution, which reassuringly holds for values of $g$ close to the critical point in sufficiently slow quenches

On the other hand, sufficiently far from the critical point inside the symmetry-broken phase, the statistics becomes identical to that of the final one up to high precision. This convergence towards the final distribution happens earlier for slower processes, in agreement with the underlying dynamics of finite-quench depth defect generation. Simulations were performed for $L_\mathrm{eff}=46$ and for $L=400$ for which it was carefully checked that for the used driving times, the effective system sizes accurately captured the low-energy dynamics.
\begin{widetext}

\begin{figure}[h]
    \centering
    \centering
    \begin{tabular}{c  c  c}
        \includegraphics[width=0.34\textwidth]{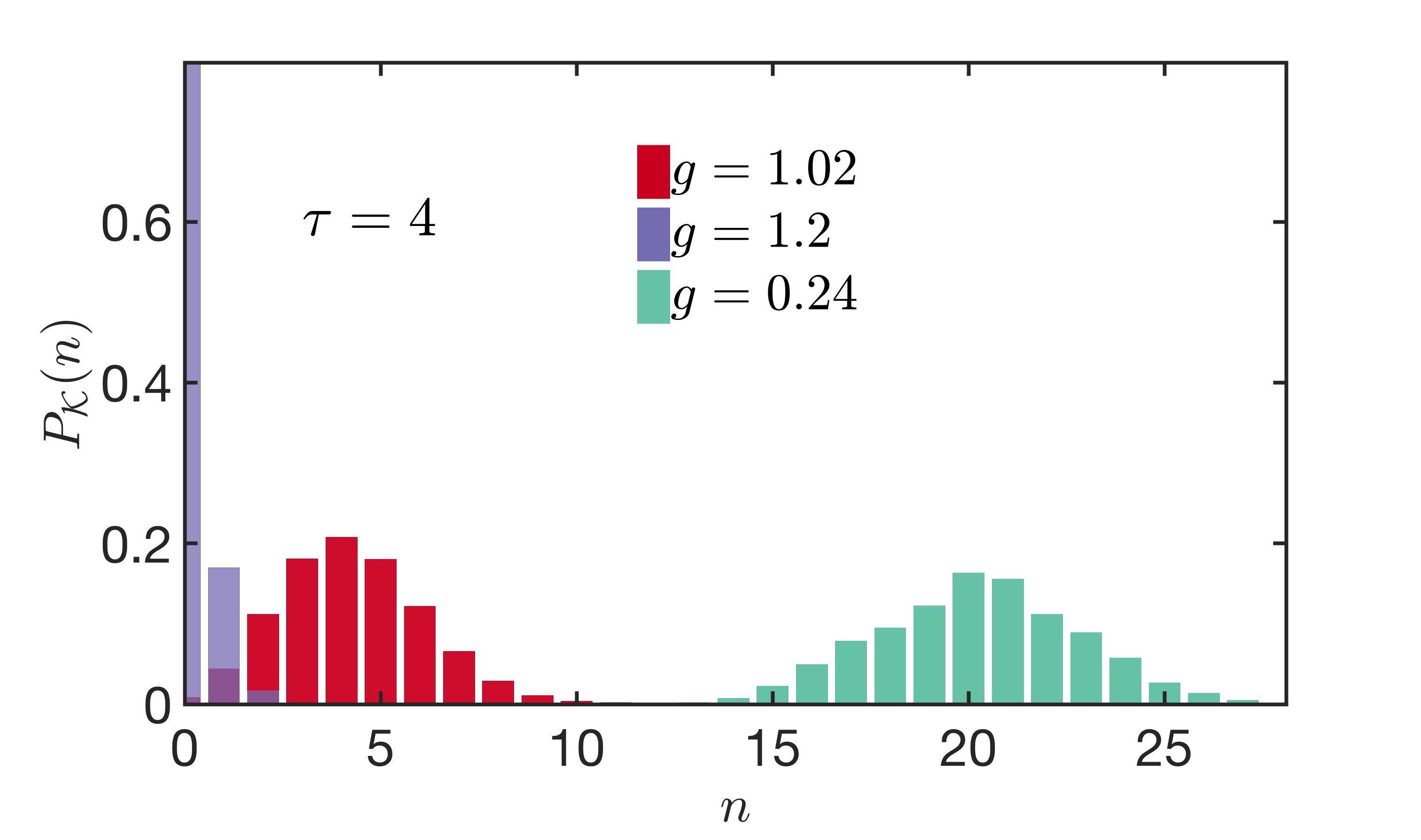} & 
        \includegraphics[width=0.34\textwidth]{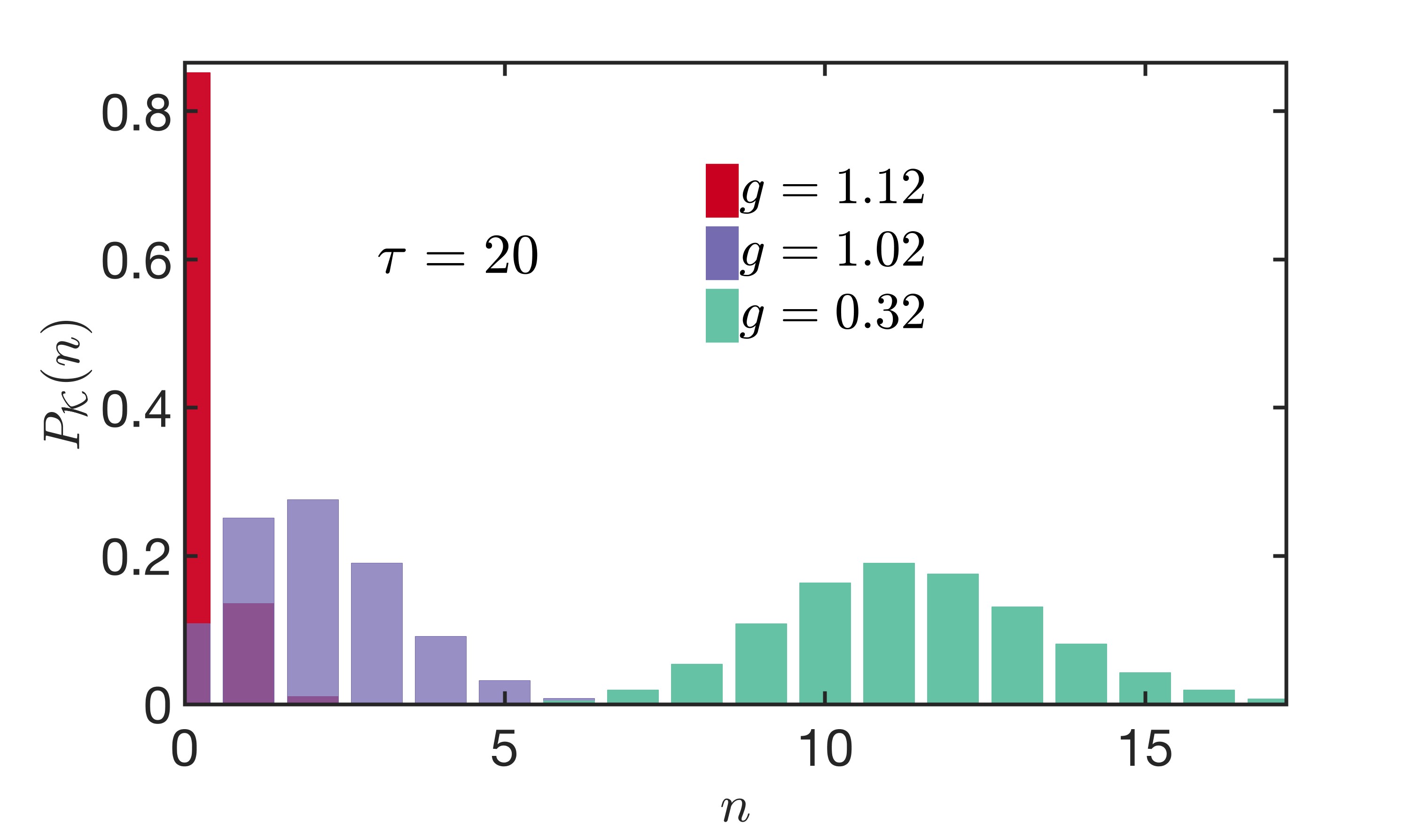} & 
        \includegraphics[width=0.34\textwidth]{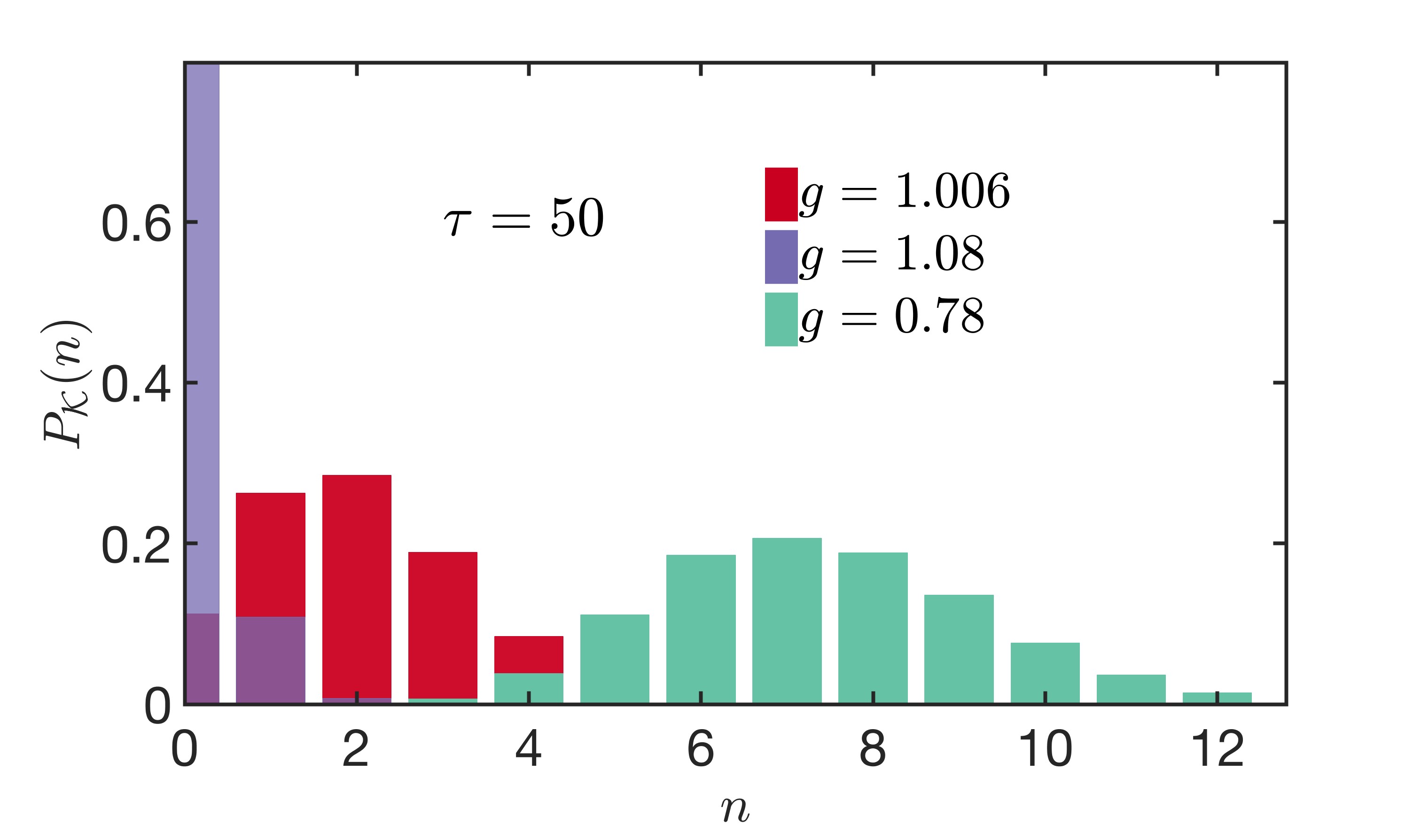} \\
    \end{tabular}
    \caption{Krylov complexity histograms for driving times $\tau=4,20,50$ for various intermediate values of $g$. In the paramagnetic phase, a peaked distribution is observed in line with adiabatic dynamics, which for faster processes starts to broaden earlier, once the critical point is approached. Inside the broken symmetry phase, the final Gaussian shape sets in earlier for slower processes $(L=400,\,L_\mathrm{eff}=46)$.}\label{fig:K_Stat}
\end{figure}

\end{widetext}
Next, we demonstrate the evolution of the second and third cumulants of the Krylov complexity. As shown in Fig.~\ref{fig: Krylov_cumulants}, the essential features remain the same as in the case of the average. A sharp growth as a universal function of $g$ is observed around the critical point. For intermediate values, stronger oscillations are traced out than for the average, which gradually vanish as the final values are reached.

\begin{figure}[h]
    \centering    \includegraphics[width=0.49\textwidth]{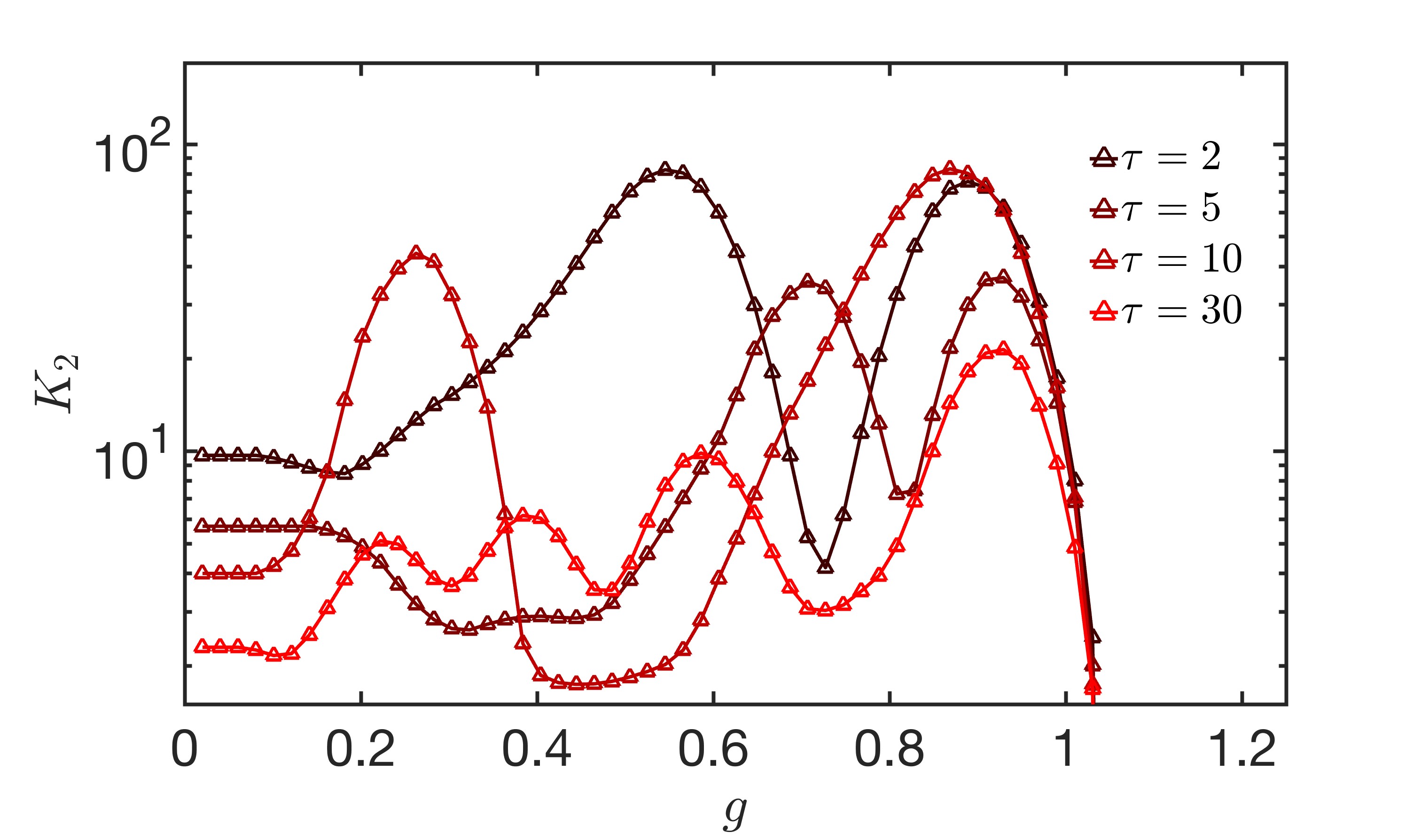}       \includegraphics[width=0.49\textwidth]{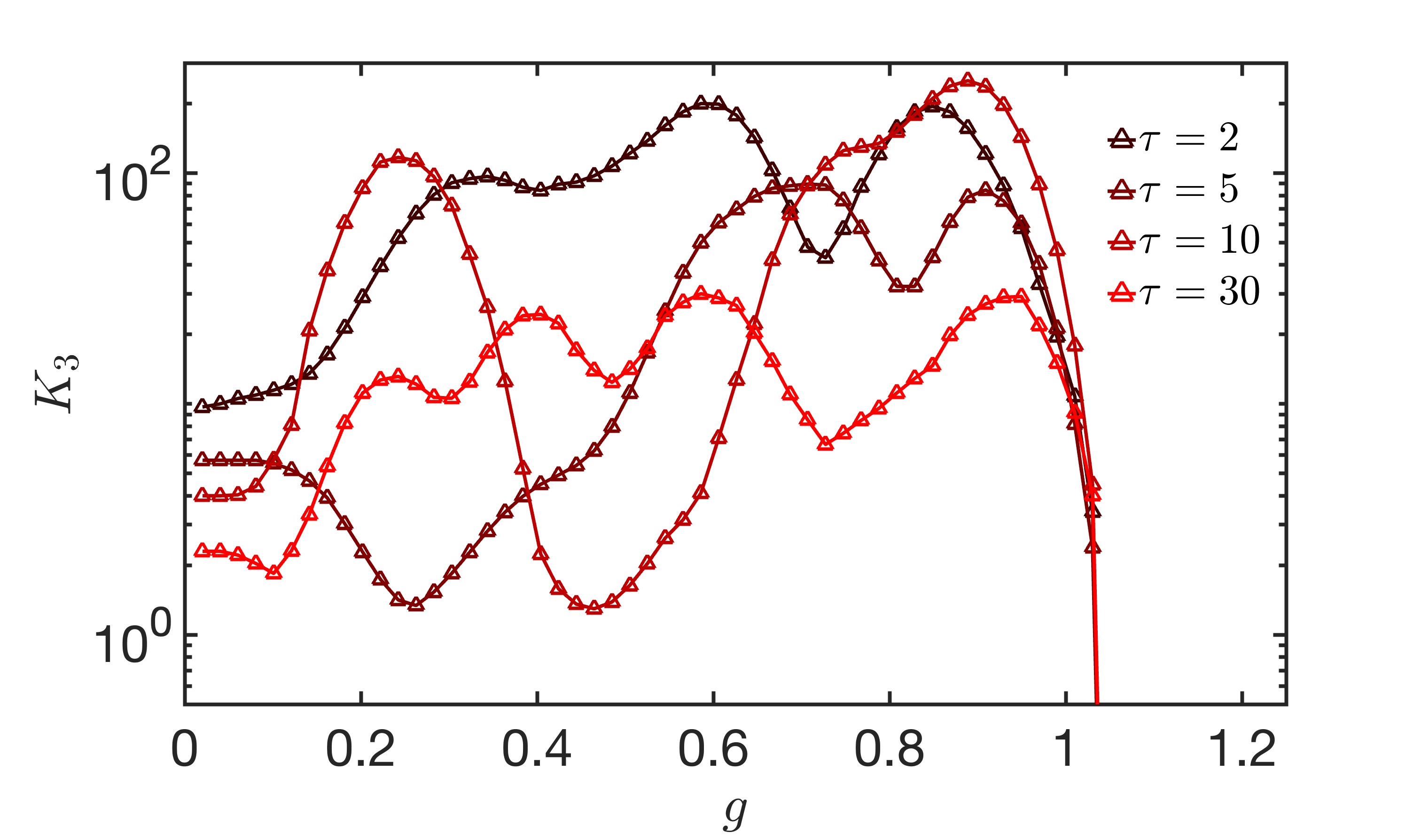}
    \caption{Variance and skewness of Krylov complexity as a function of $g$ for various values of $\tau$. Both cumulants exhibit universal growth at the critical point, followed by a highly oscillating non-universal intermediate regime. Due to the increased range of oscillations compared to the final values, the curves were plotted on a log-linear scale $(L=400,\,L_\mathrm{eff}=46)$.}\label{fig: Krylov_cumulants}
    \end{figure}

\noindent\textbf{Generalization to Free-Fermionic critical systems.}
We also provide the computational details to arrive at the general scaling behavior of the complexity statistics across second-order QPTs in free-fermion systems. Without loss of generality, assume a general quantum critical system in $d$ dimensions admitting a free fermionic representation labeled by a momentum quantum number~\cite{Polkovnikov2005}. Furthermore, assuming that the system exhibits a $d-D$ dimensional critical surface with $D$ dimensional defects, the excitation amplitudes admit a universal leading-order dependence on $\mathbf k$ and $\tau$ in the slow driving limit, $a_k(\tau)\equiv a(\mathbf k\tau^{\alpha})$. The defect density thus scales as $n_\mathrm{ex}\sim\int\mathrm d^{d-D}k\lvert a_{\mathbf k}(\tau)\rvert^2\sim\tau^{-\alpha(d-D)}$.

These dynamical critical properties are reflected in the Magnus operator in the slow driving limit as
\begin{eqnarray}
    \Omega(\tau)=\sum_{\mathbf k}\Omega(\mathbf k\tau^\alpha),
\end{eqnarray}
in the leading order, as also demonstrated by the subleading importance of the non-universal phase factors and additional $\sim O(1)$ dependencies on $\tau$ in the TFIM. Thus, the whole construction can be repeated as all multiple summations over the Magnus operator will lead to the same leading-order behavior in the Krylov states under the condition set by $n\ll L^{d-D}\tau^{-\alpha(d-D)}$, identical to $\lvert\mathbf k\rvert\ll\tau^{-\alpha}$ and by $\tau\gg L^{-\frac{1}{\alpha(d-D)}}$, i.e., with the power-law scaling regime, where transition probabilities take finite values and away from the adiabatic limit. In this regime, the same approximation applies,
\begin{eqnarray}
    &&\sum_{\mathbf k_1\neq\dots\neq \mathbf k_n}\Omega^{12}_{\lvert\mathbf k_l\rvert\tau^\alpha}\dots\Omega^{12}_{\lvert\mathbf k_n\rvert\tau^\alpha}\approx\left(\sum_{\mathbf k}\Omega^{12}_{\lvert\mathbf k\rvert\tau^\alpha}\right)^n\\
    &\sim& L^{n(d-D)}\tau^{-n(d-D)\alpha}+O\left(n\,L^{(n-1)(d-D)}\tau^{-(n-1)(d-D)\alpha}\right),\nonumber
\end{eqnarray}
along with the general derivation in App.~\cite{supp}, leading to
\begin{eqnarray}
    \lvert K_n\rangle\approx\mathcal N_n\sum_{\mathbf k_1\neq\dots\neq \mathbf k_n}\prod_{l=1}^n(\Omega^{12}_{\mathbf k_l\tau^{\alpha}},0)_{\mathbf k_l}\prod_{\mathbf k\notin\left\{\mathbf k_1,\dots,\mathbf k_n\right\}}(0,1)_{\mathbf k}^T,\qquad
\end{eqnarray}
where in the Magnus operators, in addition to the combination of $\mathbf k\tau^\alpha$, a dependence on $\tau$  appears in terms of bounded functions and in phase factors.
An additional phase factor might enter the picture; however, it will eventually vanish in the Krylov occupation numbers due to the absolute value square. As the Krylov states are given by the superposition of the $n$ times excited product states, where $n$ can range over different momentum directions perpendicular to the critical surface, the occupation probabilities and the off-diagonal matrix elements take the form
\begin{eqnarray}
    b_{n}&\sim& L^{\frac{d-D}{2}}\sqrt n\tau^{-\frac{\alpha(d-D)}{2}},\, a_{n}\sim L^{d-D}\tau^{-\alpha(d-D)},\quad\\
    \lvert\varphi_n(\tau)\rvert^2&\sim& \frac{\left(2CL^{d-D}\tau^{-\alpha(d-D)}\right)^n\,e^{-C L^{d-D}\tau^{-\alpha(d-D)}}}{n!},\qquad
\end{eqnarray}
where $C\approx\int\mathrm d^{d-D}\mathbf k \lvert a(\mathbf k)\rvert^2$ is a universality class dependent constant. Thus, all Krylov complexity cumulants are expected to be identical, scaling as
\begin{eqnarray}
    K_{\alpha,q}\approx K_{\alpha,1}=\sum_nn\lvert\varphi_n(\tau)\rvert^2\approx 2C L^{d-D}\tau^{-\alpha(d-D)}.\quad
\end{eqnarray}

\medskip

\noindent{\textbf{\textsf{Competing interests}}}\\
 The authors declare no competing interests.
 
 \smallskip

\noindent{\textbf{\textsf{Data availability}}}\\
The data supporting the findings of this study can be found via the following Zenodo link,

\smallskip

\noindent{\textbf{\textsf{Code availability}}}\\
The codes generated and used during the current study
are available from the corresponding author upon request.

 \smallskip
 
\noindent{\textbf{\textsf{Online content}}}\\
Methods, additional references, and supplementary information are available.
 
 \smallskip

\acknowledgements
\noindent{\textbf{\textsf{Acknowledgements}}}\\
 It is a pleasure to thank  Budhaditya Bhattacharjee, Kazutaka Takahashi, and Inigo Egusquiza for insightful discussions. 
This project was supported by the Luxembourg National Research Fund (FNR Grant Nos.\ 17132054 and 16434093). It has also received funding from the QuantERA II Joint Programme and co-funding from the European Union’s Horizon 2020 research and innovation programme. \\

\noindent{\textbf{\textsf{Author contributions}}}\\
 A. G. and A. D. C. performed the analytical calculations. A.G. carried out the numerical simulations. Both authors contributed to the writing of the manuscript.
 
\bibliography{references}

\newpage
\clearpage

\title{---Supplementary Material---Universal Growth of Krylov Complexity Across a Quantum Phase Transition
}
\maketitle
\onecolumngrid
\begin{center}
\textbf{\large Supplemental Material for\\
	    ``Universal Growth of Krylov Complexity Across a Quantum Phase Transition''
    }\\
    \vspace{0.5cm}
    In this Supplementary Information, detailed derivations are provided to further support the findings in the main text. We provide the computational details of the derivation of the diabatic Magnus operator. We also provide the exact derivation of the main result on the Krylov basis states in the transverse field Ising model for driving rates in the KZ scaling limit.\\
\vspace{0.5cm}
Andr\'as Grabarits and Adolfo del Campo 
\end{center}
\renewcommand{\theequation}{S\arabic{equation}}
\renewcommand{\thefigure}{S\arabic{figure}}
\renewcommand{\thetable}{S\arabic{table}}

\setcounter{equation}{0}
\setcounter{figure}{0}
\setcounter{table}{0}
\setcounter{page}{1}

\onecolumngrid

\section{Magnus operator of the TFIM}\label{app:Magnus_TFIM}
In this section, we detail how to arrive at the final form of the Magnus operator of the TFIM within the independent two-level system (TLS) representation. The time-evolution operator is obtained within the final basis, where its elements are simply the transition matrix elements of the time-evolved wave functions with the ground state and the excited state of the final Hamiltonian.
\begin{equation}
    H[0] = 
    \begin{pmatrix}
		-\cos k   	&\sin k \\
		\sin k		&\cos k
  \end{pmatrix}.
\end{equation}
It follows that
\begin{equation}
    \ket{\mathrm{GS}(0)} = 
    \begin{pmatrix}
        -\cos \frac{k}{2} \\
        \sin \frac{k}{2}
    \end{pmatrix}, \qquad
    \ket{\mathrm{ES}(0)} = 
    \begin{pmatrix}
        \sin \frac{k}{2} \\
        \cos \frac{k}{2}
    \end{pmatrix}.
\end{equation}
The exact time-evolved wave function of the TFIM in the leading order reads
\begin{eqnarray}
    \psi_{1,k} &&= \sqrt{\tau} \sin k \, e^{-\frac{\pi}{4}  \tau \sin^2 k} D_{-i \tau \sin^2 k-1} \left[ 2 e^{-3i\pi/4} \sqrt{\tau} (\cos k+t/\tau) \right], \\    
    \psi_{2,k} &&= e^{-\frac{\pi}{4}  \tau \sin^2 k+3i\pi/4} D_{-i \tau \sin^2 k} \left[2\,e^{-3i\pi/4} \sqrt{\tau} (\cos k+t/\tau) \right],
\end{eqnarray}
with $D_\nu(z)$ denoting the parabolic cylinder function. Using the compact notation, $v_k=\psi_{k,2}$, $u_k=\psi_{k,1}$ for the upcoming calculations, the time-evolution operator can be expressed as~\cite{Moessner_KZM_2024, Nowak_Dziarmaga_2022},
\begin{eqnarray}\label{eq: U_k}
    U_k(t)&&=
    \begin{pmatrix}
        v^*_k&u_k\\
        -u^*_k&v_k
    \end{pmatrix},\quad v_k\approx\sqrt{p_k}e^{i\omega_k},\,\omega_k=3\pi/4-\tau(\cos k+t/\tau)^2-\tau\sin^2k\log\left[2\sqrt\tau(\cos k+t/\tau)\right],\\
    u_k&&\approx \sqrt{1-p_k}e^{i\Phi_k},\,p_k=e^{-2\pi \tau \sin^2k},\,\Phi_k=\tau(\cos k+t/\tau)^2+\tau\sin^2k\log\left[2\sqrt\tau\left(\cos k+t/\tau\right)\right]-\mathrm{arg}\Gamma(1+i\tau\sin^2k)\nonumber\\
    &&\approx \tau(\cos k+t/\tau)^2+\tau\sin^2k\log\left[2\sqrt\tau\left(\cos k+t/\tau\right)\right]+\gamma_E\tau \sin^2k.
\end{eqnarray}
Next, the diabatic time evolution is described via the transition matrix leading to the final state starting from the instantaneous ground state. This is to account solely for nonadiabatic effects and to avoid the biasing influence of the rotating basis,
\begin{eqnarray}
    \mathbb{U}_k=\begin{pmatrix}
        -u^*_k\cos\frac{k}{2}+v^*_k\sin\frac{k}{2}&u_k\sin\frac{k}{2}+v_k\cos\frac{k}{2}\\
        -u^*_k\sin\frac{k}{2}-v^*_k\cos\frac{k}{2}&-u_k\cos\frac{k}{2}+v_k\sin\frac{k}{2}
    \end{pmatrix},\quad
    a_k\approx-u_k,\quad b_k\approx v_k,
\end{eqnarray}
which correctly reproduces the Landau-Zener result.

The Magnus operator is given via the logarithm of the general $2\times 2$ matrix of 
        \begin{eqnarray}                         
        &\Omega_k=-i\log\begin{pmatrix}
        a^*&b\\
        -b^*&a
        \end{pmatrix}=\\
    &\frac{1}{2}\begin{pmatrix}
-i\log \left( \mathrm{Re}^2\,a + (|b|^2 + \operatorname{Im}[a]^2) \right) 
-i \frac{ \operatorname{Im}[a] 
\left( 
\log \left( 2\mathrm{Re}\,a - 2i \sqrt{|b|^2 +\operatorname{Im}[a]^2} \right) 
- \log \left(2\mathrm{Re}\,a + 2i \sqrt{|b|^2+\operatorname{Im}[a]^2} \right) 
\right)}{\sqrt{|b|^2 + \operatorname{Im}[a]^2}}
\\
-\frac{b \left( 
-\log \left( 2\mathrm{Re}\,a - 2i \sqrt{|b|^2+ \operatorname{Im}[a]^2} \right) 
+ \log \left(2\mathrm{Re}\,a + 2i \sqrt{|b|^2 + \operatorname{Im}[a]^2} \right) 
\right)}{ \sqrt{|b|^2 + \operatorname{Im}[a]^2}} 
\\
-\frac{\overline{b} 
\left( 
\log \left( 2\mathrm{Re}\,a - 2i \sqrt{|b|^2+ \operatorname{Im}[a]^2} \right) 
- \log \left( 2\mathrm{Re}\,a + 2i \sqrt{|b|^2 + \operatorname{Im}[a]^2} \right) 
\right)}{ \sqrt{|b|^2 + \operatorname{Im}[a]^2}}
\\
-i\log \left( \mathrm{Re}^2\,a + (|b|^2 + \operatorname{Im}[a]^2) \right) 
+i \frac{ \operatorname{Im}[a] 
\left( 
\log \left( 2\mathrm{Re}\,a - 2 i\sqrt{|b|^2+ \operatorname{Im}[a]^2} \right) 
- \log \left( 2\mathrm{Re}\,a + 2i \sqrt{|b|^2 + \operatorname{Im}[a]^2} \right) 
\right)}{\sqrt{|b|^2 + \operatorname{Im}[a]^2}}
\end{pmatrix},
\end{eqnarray}
where the matrix elements have been enlisted in a vectorized form.
 Using $|a|^2+|b|^2=1$, the whole matrix can be parametrized as
\begin{eqnarray}
\Omega_k=\frac{\mathrm{arg}\left[\mathrm{Re}(a)+i\sqrt{1-\mathrm{Re}(a)^2}\right]}{\sqrt{1-\mathrm{Re}(a)^2}}
\begin{pmatrix}
    -\mathrm{Im}(a)&-ib\\
    ib^*&\mathrm{Im}(a)
\end{pmatrix}.
\end{eqnarray}
Next, using the leading-order expansion of the phases, one finds, for the matrix elements of the Magnus operator,
\begin{eqnarray}
    \Omega^{11}_k(t,\tau)&&\approx-\frac{\mathrm{arg}\left[\sqrt{1-p_k}\cos\Phi_k+i\sqrt{1-(1-p_k)\cos^2\Phi_k}\right]}{\sqrt{1-(1-p_k)\cos^2\Phi_k}}\sqrt{1-p_k}\sin\Phi_k,\\    
    \Omega^{12}_k(t,\tau)&&\approx i\frac{\mathrm{arg}\left[\sqrt{1-p_k}\cos\Phi_k+i\sqrt{1-(1-p_k)\cos^2\Phi_k}\right]}{\sqrt{1-(1-p_k)\cos^2\Phi_k}}\sqrt{p_k}e^{-i\omega_k},\\
    \Omega^{21}_k(t,\tau)&&=\left(\Omega^{12}_k\right)^*,\\
    \Omega^{22}_k(t,\tau)&&=-\Omega^{11}_k(t,\tau).
    \end{eqnarray}
    This result is valid for arbitrary intermediate times, as $t/\tau$ only appears as $\cos k+t/\tau$, so it does not modify any leading-order behavior with respect to $k$. Accordingly, one can further expand the phases to leading order in $k$,
    \begin{eqnarray}\label{eq:phases}
        &&\omega_k\approx-\tau(1+t/\tau)^2+\mathrm O(1),\quad\Phi_k\approx \tau(1+t/\tau)^2+\mathrm O(\tau k^2\log(\tau)),
    \end{eqnarray}
    as all terms containing $k$ will give a faster decay with $\tau$ after performing the momentum sum. As a result, the total leading order expansion reads
    \begin{eqnarray}
        \Omega^{11}_k(t,\tau)&&\approx -\frac{\mathrm{arg}\left[\left[\sqrt{p_k}\cos\left[\tau(1+t/\tau)^2\right]+i\sqrt{1-p_k\cos^2\left[\tau(1+t/\tau)^2\right]}\right]\right]}{\sqrt{1-p_k\cos^2\left[\tau(1+t/\tau)^2\right]}}\sqrt{1-p_k}\sin\left[\tau(1+t/\tau)^2\right],\\
        \Omega^{12}_k(t,\tau)&&\approx -i\frac{\mathrm{arg}\left[\left[\sqrt{p_k}\cos\left[\tau(1+t/\tau)^2\right]+i\sqrt{1-p_k\cos^2\left[\tau(1+t/\tau)^2\right]}\right]\right]}{\sqrt{1-p_k\cos^2\left[\tau(1+t/\tau)^2\right]}}\sqrt{p_k}e^{i\tau(1+t/\tau)^2}.\label{eq: Omega_12}
    \end{eqnarray}

\section{Derivation of the leading order behavior of the Krylov basis}\label{app: K_n_Magnus}
In this section, we show the steps of the proof that the $n$-th Krylov basis state is given by the superposition of the $n$ times excited product states in the leading order with respect to $L\tau^{-1/2}$. The Lanczos algorithm, generating the Krylov states, is given by
\begin{eqnarray}
    &&\lvert K_{n+1}\rangle\,b_{n+1}=\Omega\lvert K_{n}\rangle-a_{n}\lvert K_{n}\rangle-b_{n}\lvert K_{n-1}\rangle,\nonumber\\
    &&a_{n}=\langle K_{n}\lvert\Omega\rvert K_{n}\rangle,\quad b_{n}=\langle K_{n-1}\lvert\Omega\rvert K_{n}\rangle,\nonumber
\end{eqnarray}
with the initial Krylov state fixed as the instantaneous ground state, that in the final basis reads as
\begin{eqnarray}
    \lvert K_0\rangle=\prod_k(0,1)^T_k.
\end{eqnarray}
In all formulas above, the time dependence was omitted for the sake of brevity. As the maximum number of excitations in $\lvert K_n\rangle$ is $n$, in the KZ scaling regime, it holds that $n\ll L\tau^{-1/2}$. 

Next, let us study the leading-order behavior of the multiple sum over $n$ non-equal momenta. As all momentum dependence in the leading order appears via $k^2\tau$ the sum over $n$ non-equal indices can be written as
\begin{eqnarray}\label{eq: KZ_error}
    &&\sum_{k_1\neq\dots\neq k_n}\prod_{l=1}^n\Omega^{12}_{k_l}(t,\tau)\sim \left[L\tau^{-1/2}\right]^ne^{in\tau(1+t/\tau)^2}+O\left(n(L\tau^{-1/2})^{n-1}\right),\\
    &&\sum_{k_1\neq\dots\neq k_n}\prod_{l=1}^n\Omega^{12}_{k_l}(t,\tau)\Omega^{11}_{k_l}(t,\tau)\sim \left[L\tau^{-1/2}\sin\left[\tau\left(1+t/\tau\right)^2\right]\right]^n+O\left(n\left(L\tau^{-1/2}\right)^{n-1}\right),
\end{eqnarray}
where first, one neglects all constraints in the summations with a leading-order correction given by $n\sum_{k_1\neq\dots\neq k_{n-1}}$ and takes the continuum limit as $\sum_k\dots\sim L\int\mathrm dk\dots$ Note that this scaling relation is only valid for summations containing $\Omega^{12}_k$. However, as it will turn out below, this is always the case. Thus, the correction term is negligible only if $n\ll L\tau^{-1/2}$. This coincides with the condition of the KZ scaling of defects, that excitations are restricted to the regime $k\ll \tau^{-1/2}$. 

Under the assumption, proven in the following, that the $n$-th Krylov state is built from the $n$ times excited product states with weights disappearing as $p_k$ goes to zero, all Krylov states give only finite contributions for $n\ll L\tau^{-1/2}$ as the momentum is given by $k\sim n/L$. This justifies the subleading character of $n \ll \tau^{-1/2}L$.
To prove that in this limit, the leading order behavior of the $n$-th Krylov basis states is generated by the $n$ times excited states, we proceed by induction.
Consider the first step in the Lanczos algorithm, where the diagonal part of the Magnus operator $\Omega_0$ acts as the sum of $\Omega^{22}_k$ over the momenta
\begin{equation}
    \sum_k\Omega_{0,k}\lvert K_0\rangle=\left(\sum_{k}\Omega^{22}_k\right)\prod_{k}\,(0,1)^T_k.
\end{equation}
As the diagonal part acts only as a multiplying factor, the first diagonal Lanczos coefficient is equivalent to this constant, $a_0=\langle K_0\lvert \sum_{k>0}\Omega_k\rvert K_0\rangle=\sum_{k>0}\Omega^{22}_k$. As a result, the first step in the Lanczos algorithm is only generated by the off-diagonal part of the Magnus operator
\begin{eqnarray}
&&\delta\Omega=\sum_{k}
\begin{pmatrix}
    0&\Omega^{12}_k\\
    \Omega^{21}_k&0
\end{pmatrix}
=\sum_{k>0}\delta\Omega_k,\quad b_1\lvert K_1\rangle=\delta\Omega\lvert K_0\rangle=\sum_{k^\prime}(\Omega^{12}_{k^\prime},0)\prod_{k\neq k^\prime}(0,1)^T_k,
\end{eqnarray}
containing the superposition of the single excited product states. The normalization square can be evaluated via an integral approximation
\begin{eqnarray}
    b_1\approx \sqrt{L\int_0^\pi\mathrm dk\,\lvert \Omega^{12}_k\rvert^2}\sim\sqrt L\tau^{-1/4}.
\end{eqnarray}
Next, we show that in the second step, the leading-order behavior is exactly captured by the doubly excited states while the presence of the ground state and the single excited states are suppressed as $\tau^{1/4} L^{-1/2}$, which is identified as the small parameter away from the onset of adiabaticity, $\tau\ll L^2$.
First, we compute the diagonal Lanczos coefficient. These coefficients are generated only by the diagonal part of the Magnus operator; the off-diagonal parts for a given $k$ would only give non-zero results if two different product states are considered from $\lvert K_1\rangle$. In this case, however, the orthogonality of two-level states of different momenta will annul the coefficient. As a result, one can write
\begin{eqnarray}
    a_1&=&\langle K_1\lvert \sum_{k}\Omega_{0,k}\rvert K_1\rangle=b^{-2}_1\left[\sum_{k_1\neq k_2}\lvert\Omega^{12}_{k_1}\rvert^2\Omega^{22}_{k_2}+\sum_{k_1}\lvert\Omega^{12}_{k_1}\rvert^2\Omega^{11}_{k_1}\right]=\sum_{k_2}\Omega^{22}_{k_2}+b^{-2}_1\sum_{k_1}\lvert\Omega^{12}_{k_1}\rvert^2\left(\Omega^{11}_{k_1}-\Omega^{22}_{k_1}\right)\nonumber\\
    &\approx&\sum_{k_2}\Omega^{22}_{k_2}\left[1+O\left(L^{-1}\tau^{1/2}\right)\right],
\end{eqnarray}
as the second sum includes only one summation, which grows as $L\tau^{-1/2}$.
As a result, in the leading order, this main contribution from the diagonal Lanczos coefficient cancels the effects of the diagonal part of the Magnus operator,
\begin{eqnarray}
    b_2\lvert K_2\rangle&=&\left(\sum_{k}\delta\Omega_k+\sum_{k}\Omega_{0,k}-a_1\right)\lvert K_1\rangle-b_1\lvert K_0\rangle
    \nonumber\\
    &=&b^{-1}_1\Big\{\sum_{k_1\neq k_2}\left[(\Omega^{12}_{k_1},0)^T_{k_1}(\Omega^{12}_{k_2},0)^T_{k_2}+(0,\Omega^{22}_{k_2})^T_{k_2}(\Omega^{12}_{k_1},0)^T_{k_1}\right]\prod_{k\neq k_1;k_2}(0,1)^T_k\nonumber\\
    & & +\sum_{k_1}\left[(\Omega^{12}_{k_1}\Omega^{11}_{k_1},0)^T_{k_1}+(0,\Omega^{12}_{k_1}\Omega^{21}_{k_1})^T_{k_1}\right]\prod_{k\neq k_1}(0,1)^T_k-a_1\sum_{k_1}(\Omega^{12}_{k_1},0)^T_{k_1}\prod_{k_2\neq k_1}(0,1)^T_{k_2}\Big\}
    -b_1\prod_{k}(0,1)^T_k\nonumber\\
    &=&b^{-1}_1\left[\sum_{k_1\neq k_2}(\Omega^{12}_{k_1},0)^T_{k_1}(\Omega^{12}_{k_2},0)^T_{k_2}\prod_{k\neq k_1;k_2}(0,1)^T_k+\sum_{k_2}\Omega^{22}_{k_2}\sum_{k_1}(\Omega^{12}_{k_1},0)\prod_{k\neq k_1}(0,1)^T_k+\sum_{k_1}(\Omega^{11}_{k_1}-\Omega^{22}_{k_1})(\Omega^{12}_{k_1},0)\prod_{k\neq k_1}(0,1)^T_k\right]\nonumber\\
    & &\,\,+b^{-1}_1\left[\sum_{k_1}\lvert\Omega^{12}_{k_1}\rvert^2\prod_{k}(0,1)^T_k-\sum_{k_2}\Omega^{22}_{k_2}\sum_{k_1}(\Omega^{12}_{k_1},0)\prod_{k\neq k_1}(0,1)^T_k-b^{-2}_1\sum_{k_2}\lvert\Omega^{12}_{k_2}\rvert^2(\Omega^{11}_{k_2}-\Omega^{22}_{k_2})\sum_{k_1}(\Omega^{12}_{k_1},0)\prod_{k\neq k_1}(0,1)^T_k\right]\nonumber\\& &-b_1\prod_k(0,1)^T_k\nonumber\\
    &=&b^{-1}_1\left[\sum_{k_1\neq k_2}(\Omega^{12}_{k_1},0)^T_{k_1}(\Omega^{12}_{k_2},0)^T_{k_2}\prod_{k\neq k_1;k_2}(0,1)^T_k+\sum_{k_1}\left(\Omega^{11}_{k_1}-\Omega^{22}_{k_1}-b^{-2}_1\sum_{k_2}\lvert\Omega^{12}_{k_2}\rvert^2(\Omega^{11}_{k_2}-\Omega^{22}_{k_2})\right)(\Omega^{12}_{k_1},0)\prod_{k\neq k_1}(0,1)^T_k\right] \nonumber\\& &+b^{-1}_1\sum_{k_1}\lvert\Omega^{12}_{k_1}\rvert^2\prod_{k}(0,1)^T_k -b_1\prod_k(0,1)^T_k\nonumber\\
    &=&b^{-1}_1\left[\sum_{k_1\neq k_2}(\Omega^{12}_{k_1},0)^T_{k_1}(\Omega^{12}_{k_2},0)^T_{k_2}\prod_{k\neq k_1;k_2}(0,1)^T_k\right]+O(1),
    \end{eqnarray}
where we have exploited that $b^2_1=\sum_k\lvert \Omega^{12}_k\rvert^2$ and so the expression $b^{-2}_1\sum_{k_1}\lvert\Omega^{12}_{k_1}\rvert^2(\Omega^{11}_{k_1}-\Omega^{22}_{k_1})$ scales as $\sim O(1)$.
In the last equality, we also used that the dominating part of $a_1$, the double sum of the form $\sum_{k_2}\Omega^{22}_{k_2}\sum_{k_1}\Omega^{12}_{k_1}\prod_{k\neq k_1}(0,1)^T_k$ is canceled by the last term in the first line.
As a result, only single sums are left beyond the double sum over the doubly excited states. 
The order of magnitude of them are captured by the norms, scaling as $\sim b^{-2}_1 \sum_{k_1}\left\lvert \lvert\Omega^{12}_{k_1}\rvert^2\left(\Omega^{11}_{k_1}-\Omega^{22}_{k_1}\right)+\Omega^{12}_{k_1}\left(\Omega^{11}_{k_1}-\Omega^{22}_{k_1}\right)\right\rvert^2\sim L^{-1/2}\tau^{1/4}L\tau^{-1/2}\sim O(1)$. The norm of the double sum, on the other hand, expands as $\sim b^{-2}_1\sum_{k_1,k_2}\left\lvert\Omega^{12}_{k_1}\Omega^{12}_{k_2}\right\rvert^2\sim L\tau^{-1/2}$. Dominating the whole expression with a factor of $\tau^{1/4}L^{-1/2}$.
Last, even though the norm of the last ground state term equals that of the double sum term, it is exactly cancelled by the term which was transformed back to the ground state by the off-diagonal terms, $b^{-1}_1\sum_k\lvert\Omega^{12}_k\rvert^2=b_1$.

Thus, the second Krylov state is indeed generated only by the off-diagonal part of the Magnus operator, containing only doubly excited states in the leading order,
\begin{eqnarray}
    \lvert K_2\rangle=b^{-1}_1b^{-1}_2\sum_{k_1\neq k_2}(\Omega^{12}_{k_1},0)^T_{k_1}(\Omega^{12}_{k_2},0)^T_{k_2}\prod_{k\neq k_1\neq k_2}(0,1)^T_k+O(L^{-1}\tau^{1/2}),\quad b_2\sim \sqrt{2}\sqrt L\tau^{-1/4} + O(1).
\end{eqnarray}
Here, the overall normalization is given by the monic version of the Lanczos coefficients, $\tilde b^{-1}_2=b^{-1}_1\,b^{-1}_2\sim\sqrt{2!}L\tau^{-1/2},\quad b^{2}_1\sim b^{2}_2/2\sim L\tau^{-1/2}$.

Having verified that the construction works for both the trivial $\lvert K_1\rangle$ state and the first non-trivial $\lvert K_2\rangle$ state, one can prove by induction that the leading order behavior holds for all $\lvert K_n\rangle$, $n<L/2$. To this end, let us assume that it holds for $n$, that
\begin{eqnarray}
    \lvert K_n\rangle&=&\tilde b^{-1}_n\sum_{k_1\neq\dots\neq k_n}(\Omega^{12}_{k_1},0)^T_{k_1}\dots(\Omega^{12}_{k_n},0)^T_{k_n}\prod_{k\notin\{ k_1;\dots;k_n\}}(0,1)^T_k,\\
    \tilde b^{2}_n&=&n!\sum_{k_1\neq\dots\neq k_n}\lvert\Omega^{12}_{k_1}\dots\Omega^{12}_{k_n}\rvert^2=\prod_{l=1}^nb^{2}_l,\,b^{2}_l\sim L\tau^{-1/2}\,l\,.\label{eq: title b_n}
\end{eqnarray}
Now, in the next step, similar features can be observed as in the case of $\lvert K_2\rangle$. First, we compute again the $n$-th diagonal Lanczos coefficient up to the leading order, which is only generated by the diagonal part of the Magnus operator, $\sum_k\Omega_{0,k}$,
    \begin{eqnarray}\label{eq: a_n}
        a_n&&=\langle K_n\lvert \sum_k\delta\Omega_k+\Omega_{0,k}\rvert K_n\rangle=\langle K_n\lvert \sum_k\Omega_{0,k}\rvert K_n\rangle\nonumber\\
        &&=n!\,\tilde b^{-2}_n\left[\sum_{k_1\neq k_2\neq\dots\neq k_n}\lvert\Omega^{12}_{k_1}\Omega^{12}_{k_2}\dots\Omega^{12}_{k_n}\rvert^2\sum_{k\notin\{k_1\neq\dots\neq k_n\}}\Omega^{22}_{k}+\sum_{k_1\neq k_2\neq\dots\neq k_n}\lvert\Omega^{12}_{k_1}\Omega^{12}_{k_2}\dots\Omega^{12}_{k_n}\rvert^2\sum_{l=1}^n\Omega^{11}_{k_l}\right]\nonumber\\
        &&=n!\,\tilde b^{-2}_n\left[\sum_k\Omega^{22}_k\sum_{k_1\neq k_2\neq\dots\neq k_n}\lvert\Omega^{12}_{k_1}\Omega^{12}_{k_2}\dots\Omega^{12}_{k_n}\rvert^2+\sum_{l=1}^n\sum_{k_1\neq k_2\neq\dots\neq k_n}\lvert\Omega^{12}_{k_1}\Omega^{12}_{k_2}\dots\Omega^{12}_{k_n}\rvert^2\left(\Omega^{11}_{k_l}-\Omega^{22}_{k_l}\right)\right]\nonumber\\
        &&\approx\sum_k\Omega^{22}_k+n!\tilde b^{-2}_n\,n\sum_{k_1\neq\dots k_n}\lvert\Omega^{12}_{k_1}\dots\Omega^{12}_{k_n}\rvert^2(\Omega^{11}_{k_1}-\Omega^{22}_{k_1}),
    \end{eqnarray}
by the definition of $\tilde b_n$, Eq.~\eqref{eq: title b_n}.
In contrast to the $n=1$ case, here the second term cannot generally be neglected due to the additional parameter of $n$. However, the first sum scales as $L\tau^{-1/2}$, which dominates over $n$ within the KZ scaling regime as 
\begin{eqnarray}
    L\tau^{-1/2}\gg n\Rightarrow n/L\sim k\ll \tau^{-1/2}.
\end{eqnarray}
Next, we write the $n+1$-th step in the Lanczos algorithm, where for convenience, we write out the pure product states with the resulting monic coefficient, $\tilde b_n$ put to the l.h.s.,

\begin{eqnarray}
    \tilde b_nb_{n+1}\lvert K_{n+1}\rangle&=&\tilde b_n\left(\sum_{k}\delta\Omega_k+\sum_{k}\Omega_{0,k}-a_n\right)\lvert K_n\rangle-b_n\tilde b_n\lvert K_{n-1}\rangle\\
    &=&\sum_{k_1\neq\dots\neq k_n\neq k_{n+1}}\prod_{k\in\left\{k_1\neq\dots\neq k_n\neq k_{n+1}\right\}}(\Omega^{12}_{k},0)^T_{k}\prod_{k\notin\{k_1\neq\dots\neq k_n\neq k_{n+1}\}}(0,1)^T_k\nonumber\\
    &&+\sum_{k_1\neq\dots\neq k_n}\prod_{k\in\left\{k_1\neq\dots\neq k_n\right\}}(\Omega^{12}_{k},0)^T_{k}\sum_{m=1}^n\Omega^{11}_{k_m}\prod_{k\notin\{k_1\neq\dots\neq k_n\}}(0,1)^T_k\nonumber\\
    &&+\sum_{k_1\neq\dots\neq k_n}\prod_{k\in\left\{k_1\neq\dots\neq k_n\right\}}(\Omega^{12}_{k},0)^T_{k}\sum_{k\notin\{k_1\neq\dots\neq k_n\}}\Omega^{22}_{k}\prod_{k\notin\{k_1\neq\dots\neq k_n\}}(0,1)^T_k\nonumber\\
    &&-a_n\sum_{k_1\neq\dots\neq k_n}(\Omega^{12}_{k_1},0)^T_{k_1}\dots(\Omega^{12}_{k_n},0)^T_{k_n}\prod_{k\notin\{ k_1;\dots;k_n\}}(0,1)^T_k\nonumber\\
    &&+\sum_{k_1\neq\dots\neq k_{n}}\prod_{k\in\{k_1\neq\dots\neq k_{n-1}\}}\!\!(\Omega^{12}_k,0)^T_k(0,\lvert\Omega^{12}_{k_n}\rvert^2)^T_{k_n}\prod_{k\notin\{k_1\neq\dots\neq k_{n}\}}\!\!\!\!\!\!\!\!(0,1)^T_k\nonumber\\
    &&- b^2_n\sum_{k_1\neq\dots\neq k_{n-1}}\prod_{k\in\{k_1,\dots, k_{n-1}\}}\!\!\!\!\!\!(\Omega^{12}_{k},0)^T_k\prod_{k\notin\{k_1\dots, k_{n-1}\}}\!\!\!\!\!\!(0,1)^T_k.\nonumber    
\end{eqnarray}
The second line describes the $n+1$ times excited states, while the third and fourth lines correspond to the cases where the diagonal elements acted on the excited states and the ground states in the given $k$ modes, respectively. The fifth line gives the contributions of the diagonal Lanczos coefficients, while the last line corresponds to the case where the off-diagonal Magnus operator reduces the number of excited states by one, and the last part of the Lanczos algorithm, the coefficient of which was computed as $\tilde b_nb_n\tilde b^{-1}_{n-1}\approx b^2_n$ in the leading order as given in Eq.~\eqref{eq: title b_n}.

By this, the last line is exactly canceled in the leading order, leaving no contributions coming from $n-1$ times excited states. To see this, realize that $\tilde b^2_n=n!\sum_{k_1\neq\dots\neq k_n}\lvert\Omega^{12}_{k_1}\dots\Omega^{12}_{k_n}\rvert^2$ and $b_n=\tilde b_n/\tilde b_{n-1}$ (Eq.~\eqref{eq: title b_n}), leading to $b^2_n\approx n\sum_k\lvert\Omega^{12}_k\rvert^2$. At the same time, in the leading order, the first term in the last line reads

\begin{eqnarray}
    &&\sum_{k_1\neq\dots\neq k_{n}}\prod_{k\in\{k_1\neq\dots\neq k_{n-1}\}}(\Omega^{12}_k,0)^T_k(0,\lvert\Omega^{12}_{k_n}\rvert^2)^T_{k_n}\prod_{k\notin\{k_1\neq\dots\neq k_{n}\}}(0,1)^T_k\\
    &&\approx n\sum_{k}\lvert\Omega^{12}_{k}\rvert^2\,\sum_{k_1\neq\dots\neq k_{n-1}}\prod_{k\in\{k_1\neq\dots\neq k_{n-1}\}}(\Omega^{12}_k,0)^T_k\prod_{k\notin\{k_1\neq\dots\neq k_{n}\}}(0,1)^T_k,\nonumber
\end{eqnarray}
verifying the cancellation. Note that the correction terms in both steps scale with the already fixed error term $\sim n\sqrt\tau/L\ll1$ in the KZ scaling regime, Eq.~\eqref{eq: KZ_error}.

As a result, the remaining expression to be addressed, where for completeness, we also write the full form of $a_n$, Eq.~\eqref{eq: a_n}, 
\begin{eqnarray}
    &&\sum_{k_1\neq\dots\neq k_n\neq k_{n+1}}\prod_{k\in\left\{k_1\neq\dots\neq k_n\neq k_{n+1}\right\}}(\Omega^{12}_{k},0)^T_{k}\prod_{k\notin\{k_1\neq\dots\neq k_n\neq k_{n+1}\}}(0,1)^T_k\\
    &&+\sum_{k_1\neq\dots\neq k_n}\left(\sum_{m=1}^n\Omega^{11}_{k_m}+\sum_{k\notin\{k_1\neq\dots\neq k_n\}}\Omega^{22}_{k}\right)\prod_{k\in\left\{k_1\neq\dots\neq k_n\right\}}(\Omega^{12}_{k},0)^T_{k}\prod_{k\notin\{k_1\neq\dots\neq k_n\}}(0,1)^T_k\nonumber\\
    &&-n!\,\tilde b^{-2}_n\left[\sum_k\Omega^{22}_k\sum_{k_1\neq k_2\neq\dots\neq k_n}\lvert\Omega^{12}_{k_1}\Omega^{12}_{k_2}\dots\Omega^{12}_{k_n}\rvert^2+\sum_{l=1}^n\sum_{k_1\neq k_2\neq\dots\neq k_n}\lvert\Omega^{12}_{k_1}\Omega^{12}_{k_2}\dots\Omega^{12}_{k_n}\rvert^2\left(\Omega^{11}_{k_l}-\Omega^{22}_{k_l}\right)\right]\nonumber\\
    &&\times\sum_{k_1\neq\dots\neq k_n}\prod_{k\in\{k_1,\dots,k_n\}}(\Omega^{12}_{k},0)^T_{k}\prod_{k\notin\{ k_1;\dots;k_n\}}(0,1)^T_k\nonumber\\
    =&&\sum_{k_1\neq\dots\neq k_n\neq k_{n+1}}\prod_{k\in\left\{k_1\neq\dots\neq k_n\neq k_{n+1}\right\}}(\Omega^{12}_{k},0)^T_{k}\prod_{k\notin\{k_1\neq\dots\neq k_n\neq k_{n+1}\}}(0,1)^T_k\nonumber\\
    &&+\sum_k\Omega^{22}_k\left[1-n!\tilde b^{-2}_n\sum_{k_1\neq\dots\neq k_n}\lvert\Omega^{12}_{k_1}\dots \Omega^{12}_{k_n}\rvert^2\right]\sum_{k_1\neq\dots\neq k_n}\prod_{k\in\left\{k_1\neq\dots\neq k_n\right\}}(\Omega^{12}_{k},0)^T_{k}\prod_{k\notin\{k_1\neq\dots\neq k_n\}}(0,1)^T_k\nonumber\\
    &&+\sum_{k_1\neq\dots\neq k_n}\sum_{m=1}^n\left[(\Omega^{11}_{k_m}-\Omega^{22}_{k_m})-n!\tilde b^{-2}_n\sum_{k^\prime_1\neq\dots\neq k^\prime_n}\lvert\Omega^{12}_{k^\prime_1}\dots\Omega^{12}_{k^\prime_n}\rvert^2(\Omega^{11}_{k^\prime_m}-\Omega^{22}_{k^\prime_m})\right]\prod_{k\in\left\{k_1\neq\dots\neq k_n\right\}}\!\!\!\!\!\!(\Omega^{12}_{k},0)^T_{k}\prod_{k\notin\{k_1\neq\dots\neq k_n\}}\!\!\!\!\!\!(0,1)^T_k.\nonumber
\end{eqnarray}

Here, in the second line, we just expressed the diagonal Lanczos coefficients, $a_n$, from Eq.~\eqref{eq: a_n}. In the second part, we reordered the $n$ times excited product states.
The sixth line cancels exactly in the leading order, as $\tilde b^2_n=n!\sum_{k_1\neq\dots\neq k_n}\lvert\Omega^{12}_{k_1}\dots\Omega^{12}_{k_n}\rvert^2+O\left(n!\,n(L\tau^{-1/2})^{n-1}\right)$. Thus, one needs to handle only the last line to verify that the dominant term is the purely $n+1$ time excited product states. The second non-equal index sum can be evaluated in the leading order as

\begin{eqnarray}
    n!\sum_{m=1}^n\sum_{k^\prime_1\neq\dots\neq k^\prime_n}\lvert\Omega^{12}_{k^\prime_1}\dots\Omega^{12}_{k^\prime_n}\rvert^2(\Omega^{11}_{k^\prime_m}-\Omega^{22}_{k^\prime_m}) &\approx& n\,n!\left(\sum_{k}\lvert\Omega^{12}_k\rvert^2\right)^{n-1}\sum_k\lvert\Omega^{12}_k\rvert^22\mathrm{Re}\,\Omega^{11}_k\nonumber\\
    &\approx& n\,\tilde b^2_n\sin\left[\tau(1+t/\tau)^2\right]\equiv n\,c\,\tilde b^2_n.
\end{eqnarray}
Using the shorthand notation, $\Omega^r_k\equiv2\mathrm{Re}\,\Omega^{11}_k$, the $n$ times excited states and their norm square read
\begin{eqnarray}
    &&n!\sum_{k_1<\dots< k_n}\sum_{m=1}^n(\Omega^{r}_{k_m}-c)\prod_{k\in\left\{k_1<\dots< k_n\right\}}(\Omega^{12}_{k},0)^T_{k}\prod_{k\notin\{k_1<\dots< k_n\}}(0,1)^T_k\\
    &&\Rightarrow n!\sum_{m,m^\prime=1}^n\sum_{k_1\neq \dots\neq k_n}\lvert\Omega^{12}_{k_1}\dots\Omega^{12}_{k_n}\rvert^2(\Omega^r_{k_m}-c)(\Omega^r_{k_{m^\prime}}-c)=n!\sum_{m=1}^n\sum_{k_1\neq \dots\neq k_n}\lvert\Omega^{12}_{k_1}\dots\Omega^{12}_{k_n}\rvert^2(\Omega^r_{k_m}-c)^2\sim n\,n! L^{n}\tau^{-n/2}.\nonumber
\end{eqnarray}
In the leading order, the above expression is only non-zero for $m=m^\prime$ by the definition of the constant $c$, just canceling sums as $\sum_k\lvert\Omega^{12}_k\rvert^2(\Omega^{r}_k-c)=0$. As a result, the norm of the $n$-times excited states grows as $\sim n\,n!L^n\tau^{-n/2}$. At the same time, the norm of the $n+1$ times excited states scales as
\begin{eqnarray}
    (n+1)!\sum_{k_1\neq\dots\neq k_{n+1}}\lvert\Omega^{12}_{k_1}\dots\Omega^{12}_{k_{n+1}}\rvert^2\sim (n+1)!L^{n+1}\tau^{-(n+1)/2},
\end{eqnarray}
verifying that the norm square of the $n+1$ times excited states dominates over $n$ times excited ones with a factor of $L\,\tau^{-1/2}$.

\end{document}